\documentclass[12pt]{iopart}
\pdfoutput=1
\usepackage{amsfonts}
\usepackage{graphicx}

\usepackage{xcolor}
\usepackage{bm}
\usepackage{subfigure}
\usepackage{textcomp}
\usepackage{cite}
\usepackage{float}
\usepackage{soul}
\usepackage{stackrel}
\usepackage{ulem}
\usepackage{multirow}
\usepackage{mathrsfs}

\usepackage{stackengine}

\usepackage{booktabs}

\newcommand\floor[1]{ \left \lfloor#1 \right \rfloor}

\definecolor{dgreen}{rgb}{0,0.7,0}

\expandafter\let\csname equation*\endcsname\relax
\expandafter\let\csname endequation*\endcsname\relax
\usepackage{amsmath,amssymb}

\pdfoutput=1
\usepackage{esint}

\usepackage{hyperref}
\usepackage{color}
\definecolor{dvo}{rgb}{0.7,0.2,0.2}
\definecolor{dgreen}{rgb}{0,0.7,0}

\begin{document}

\title[]{Mean area of the convex hull of a run and tumble particle in two dimensions}

\author{Prashant Singh$^1$, Anupam Kundu$^1$, Satya N. Majumdar$^2$ and Hendrik Schawe$^3$}

\address{International Centre for Theoretical Sciences, Tata Institute of Fundamental ~Research, Bengaluru 560089, India $^1$}
\address{LPTMS, CNRS, Univ. Paris-Sud, Universit\'e Paris-Saclay, 91405, Orsay, France $^2$}
\address{Laboratoire de Physique Th\'{e}orique et Mod\'{e}lisation, UMR-8089 CNRS, CY
Cergy Paris Universit\'{e}, France $^3$}
\ead{prashant.singh@icts.res.in}
\ead{anupam.kundu@icts.res.in}
\vspace{10pt}

\begin{abstract}
We investigate the statistics of the convex hull for a single run-and-tumble particle in two dimensions. Run-and-tumble particle (RTP), also known as persistent random walker, has gained significant interest in the recent years due to its biological application in modelling the motion of bacteria. We consider two different statistical ensembles depending on whether (i) the total number of tumbles $n$ or (ii) the total observation time $t$ is kept fixed. Benchmarking the results on perimeter, we study the statistical properties of the area of the convex hull for RTP. Exploiting the connections to extreme value statistics, we obtain exact analytical expressions for the mean area for both ensembles. For fixed-$t$ ensemble, we show that the mean possesses a scaling form in $\gamma t$ (with $\gamma$ being the tumbling rate) and the corresponding scaling function is exactly computed. Interestingly, we find that it exhibits crossover from $\sim t^3$ scaling at small times $\left( t \ll \gamma ^{-1} \right)$ to $\sim t$ scaling at large times $\left( t \gg \gamma ^{-1} \right)$. On the other hand, for fixed-$n$ ensemble, the mean expectedly grows linearly with $n$ for $n \gg 1$. All our analytical findings are supported with numerical simulations.


\end{abstract}

\section{Introduction}
\label{intro}
Active matter refers to a class of driven non-equilibrium systems that transduces systematic movement out of the supplied energy. Contrary to the boundary-driven systems, the energy is exchanged at the local scale which endows the constituent particles with self-propulsion {\cite{Ramaswamy 2010, Ramaswamy 2017, Bechinger 2016}}. As a result, the dynamics of these systems break time-reversal symmetry and thus, violate the detailed balance. Recently, the self-propulsion (or ``active" nature) has been harnessed to produce useful work for potential therapeutic applications in various diseases like cancer and heart disease {\cite{Santiago2018, Ghosh2020}}. Furthermore, they reveal a plethora of complex features like clustering {\cite{Redner 2013, Bricard 2013}}, flocking {\cite{Kumar2014, Toner2005}}, motility induced phase separation {\cite{Cates 2015,Gonnella 2015,Partridge 2019,Caprini2020 }}, non-existence of equation of states for pressure {\cite{Solon 2015}} and so on. Going beyond the theoretical studies, the dynamics of active particles has been realised in many experiments based on different phoretic effects {\cite{Howse2007, Jiang2010}}.

\begin{figure}[t]
\includegraphics[scale=0.45]{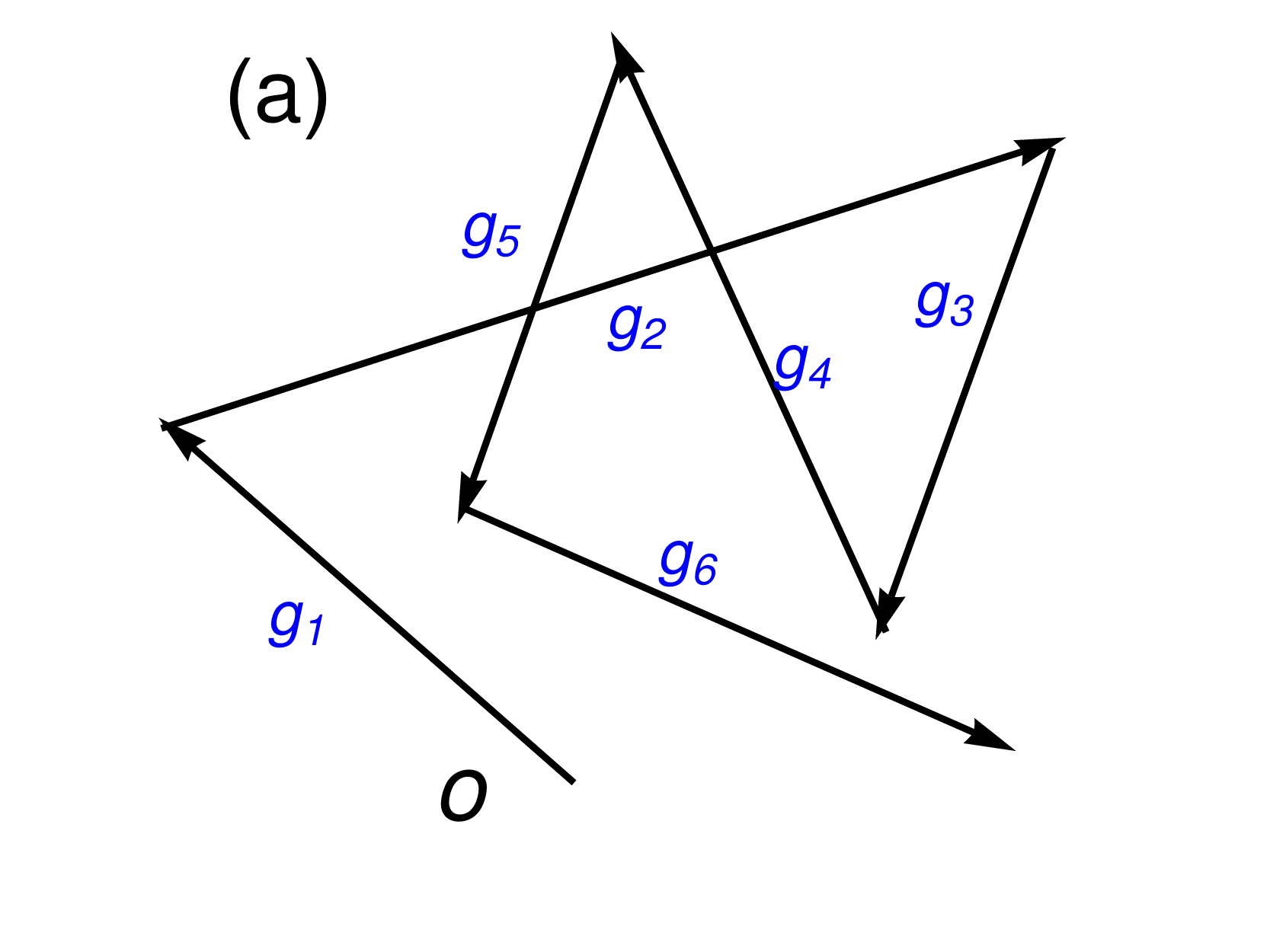}
\includegraphics[scale=0.45]{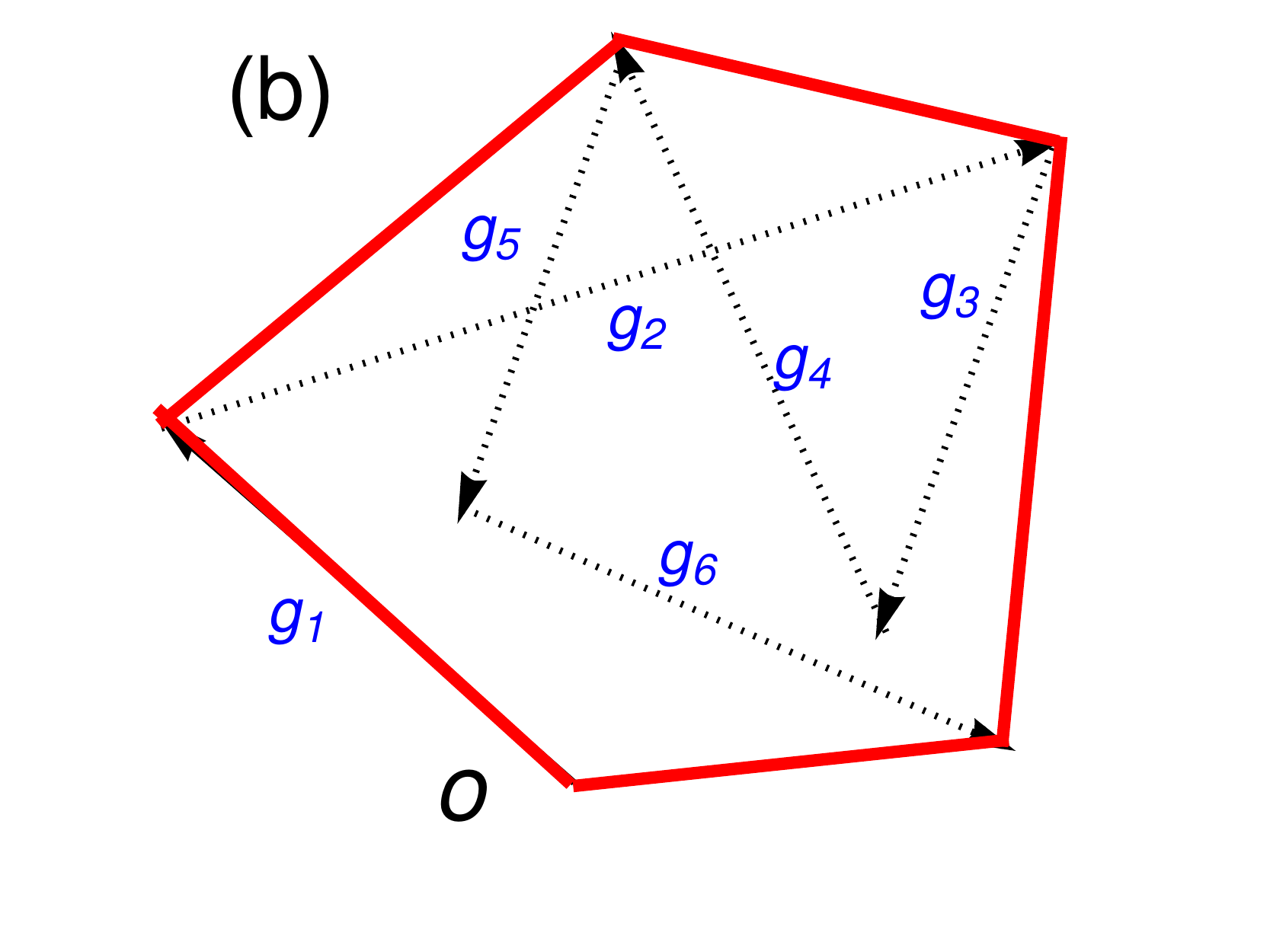}
\centering
\caption{(a) Schematic representation of a typical trajectory of a RTP in two dimensions with total number of runs $n=6$. The RTP moves in a series of runs interspersed by instantaneous tumbles that occur after random times with rate $\gamma$. (b) This figure shows the convex hull (red polygon) for the trajectory on the left. }    
\label{fig-trajectory}
\end{figure}

Run and tumble particle (RTP) has emerged as a quintessential model in mimicking the dynamics of active particles. Previously known in the random walk literature as persistent Brownian motion {\cite{Masoliver1993, Weiss2002}}, the RTP {motion} has recently been quite extensively studied due to its biological application in modelling the motion of bacteria like E Coli {\cite{Berg2003, Tailleur2008,Cates 2015}}. In this model, , the particle moves ballistically along a certain direction with constant speed $v_0~(\geq 0)$ till a random time {duration} $\tau $ drawn from the exponential distribution with constant rate $\gamma$, i.e. {from the distribution} $\rho(\tau) = \gamma e^{-\gamma \tau}$. The event of ballistic motion is referred to as a `run'. After the random time {duration} $\tau_1$, the particle undergoes `tumbling' in which it chooses a new direction uniformly. {In RTP models, the tumble events are assumed to occur instantaneously. After the tumble event,} it runs along this new direction with constant speed $v_0$ for another {random time duration $\tau_2$, again} drawn independently from $\rho(\tau) = \gamma e^{-\gamma \tau}$ . In this way, the RTP moves in a series of runs interspersed by instantaneous tumbles that occur after random time {durations} with rate $\gamma$. A schematic representation of the trajectory is shown in Figure \ref{fig-trajectory}(a). Over the recent few years, this model has been substantially studied and a variety of results are known. Examples include - position distribution in free space as well as in confining potential {\cite{Kanaya_2018,Dhar2019, Demaerel2018,Evans2018, SinghSabhapandit2020, Santra2020,Doussal2020}}, condensation transition \cite{Mori2021cond, Gradenigo2019, Mori2021}, persistent properties {\cite{Angelani2014,Mori2019, Bruyn2021}}, extremal properties {\cite{Singh2019,MoriDoussal2020}}, path functionals {\cite{Singh2019,Singh2021}}, current fluctuations {\cite{Banerjee2020}}, interacting multiple RTPs {\cite{Doussal2020, Slowman2016,SinghKundu2021,DoussalMajumdar2021, Doussal2019noncross}}, etc.

{In this paper}, we are interested in the statistics of convex hull for the RTP in two dimensions. Consider a set of points $( \vec{r}_1,\vec{r}_2,...,\vec{r}_N )$ in two dimensions. For simplicity, one can think of them as position of a particle at various instances of time. Then, convex hull refers to the unique smallest convex polygon that encloses all these points {\cite{Letac1980, Hug2013}} [see Figure \ref{fig-trajectory}(b)]. Now, for a stochastic process, the set $( \vec{r}_1,\vec{r}_2,...,\vec{r}_N )$ varies over realisations which implies that the convex hull  is also different for different realisations. One is, then, interested in the statistical properties of this random convex hull. In ecology, convex hull has been used in estimating the extent over which the animals move during foraging or other activities  {\cite{Wortan1995}}. Clearly, this knowledge is useful in designing and demarcating the geographical territory for them. The properties of convex hull have been of prime interest in the mathematics literature also {\cite{Kac1954, Spitzer1956, Snyder1993,Kabluchko2017,KabluchkoVysotsky2017}}. In physics, the mean area and mean perimeter of the convex hull are found to be related to the subject of extreme value statistics  {\cite{MajumdarFurling2010}}. Exploiting this connection, the mean area and mean perimeter have been studied for a variety of processes like Brownian motion {\cite{Furling2009}}, random acceleration {\cite{Reymbaut2011}}, diffusion with resetting {\cite{MajumdarMoriSchawe2021}}, random walk and its generalisations {\cite{Grebenkov2017, Claussen 2015, Kampf2012,Lukovic2013,Dumonteil2013 ,Schawe2018,Schawe2019, Chupeau2015}}. Extensions of these studies to higher dimensions and multi-particle case have also been considered {\cite{Eldan2014,Kabluchko2016, Schawe2017, MajumdarFurling2010}}. Going beyond the mean values, the entire distributions of the area and perimeter have also been studied using sophisticated numerical techniques {\cite{Claussen 2015,Schawe2018,Schawe2019}}. We refer to {\cite{MajumdarFurling2010}} for a review on the convex hull {problem}. 

{Recently in {\cite{Hartmann2020}}, the mean perimeter of convex hull for RTP in a plane was exactly computed for {the} two different ensembles - (i) fixed number of tumbles $n$ and (ii) fixed observation time $t$ (discussed later). Here, we go beyond this work to investigate the statistics of the area of convex hull for two dimensional RTP. For both ensembles, we  compute the mean area exactly. We verify our analytical results numerically and, also study the variance and the distribution of the area numerically.} 

{The paper is organised as follows: In Sec. \ref{model}, we introduce the model and summarize the main results of our paper. Sec. \ref{mean-area-exact} contains a brief discussion on convex hull problems for general $2$-$d$ stochastic processes. Analytic calculations for mean area are presented in Sec. \ref{mean-area-fixed-n-cal} for fixed-$n$ ensemble and in Sec. \ref{mean-area-fixed-t-cal} for fixed-$t$ ensemble. We devote Sec. \ref{numerical dist} for the numerical study of probability distribution of the area which is followed by the conclusion in Sec. \ref{conclusion}. }

\section{Model and summary of the results}
\label{model}
We consider a RTP moving on a plane. Starting from the origin, the particle chooses an angle $\phi _ 1$ (measured with respect to the $x$-axis) uniformly from $[0, 2 \pi]$ and moves ballistically in that direction with a speed $v_0$. The ballistic motion, referred to as `run', persists along $\phi _1$ for a random time $\tau _1$ drawn from exponential distribution $\rho(\tau) = \gamma e^{-\gamma \tau}$ with constant rate $\gamma$. After this, the particle `tumbles' {instantaneously} in which {event} it chooses a new direction $\phi _2$ uniformly from $[0, 2 \pi]$. Then, it performs another run for random time $\tau _2$ {again} drawn independently from $\rho(\tau) = \gamma e^{-\gamma \tau}$. The motion continues in the form of the ballistic \textit{runs} interspersed by the instantaneous \textit{tumbles} that occur after random times drawn independently from exponential distribution. Let us focus on the $i$-th run along the direction $\phi _i$. Denoting the {displacement (position increment)} during $i$-th interval by $(x_i, y_i)$, we have
\begin{align}
x_i = v_0 \tau _i \cos (\phi _i), \label{x-cord-eq-1} \\
y_i = v_0 \tau _i \sin (\phi _i), \label{y-cord-eq-1} 
\end{align} 
where $\tau _i$ is the time till which the $i$-th run lasts. The position of the particle $(X_i, Y_i)$ after $i$-th run can be written in terms of $(x_i, y_i)$ as
\begin{align}
X_{i} &= X_{i-1}+x_i, \label{x-cord-eq-2}\\
Y_{i} &= Y_{i-1}+y_i, \label{y-cord-eq-2}
\end{align}
where $i=1,2,...$ and we assume $(X_0, Y_0) = (0,0)$. {As mentioned earlier,} we consider the motion of the particle in two different ensembles with - {(i)} fixed $n$ and {(ii)} fixed $t$. In case {(i)}, the particle undergoes a fixed number of runs (say $n$) and we stop the process after $n$ runs have taken place. The total observation time $t$ will fluctuate for different realisations. Moreover, we consider the starting point as a tumble which makes the number of runs equal to the number of tumblings and $n \geq 1$. On the other hand, for ensemble {(ii)}, we fix the total observation time $t$ and therefore, the number of tumblings $n$ fluctuates for different realisations.

{For these two ensembles, we look at the statistical properties of convex hull. Recently, the mean perimeter of convex hull for this model was computed exactly  for the two ensembles and the distribution for the perimeter was numerically studied {\cite{Hartmann2020}}. Here, we investigate the statistics of area of the convex hull both analytically and numerically. Using connection to the extreme value statistics developed in {\cite{Furling2009, MajumdarFurling2010}}, we  compute the mean area in the two ensembles exactly. Next, we also investigate the variance and distribution of the typical fluctuations of the area. Our main results are summarised below:}
\begin{enumerate}
\item For fixed $n$ ensemble, we find that the mean area $\langle A_n \rangle$ is given by
\begin{align}
\langle A_n \rangle = \frac{v_0^2}{2 \gamma ^2} \mathcal{S}_n,~~n>1
\label{fix-n-area}
\end{align}
where the term $\mathcal{S}_n$ is given by 
\begin{align}
\mathcal{S}_n& =\frac{2+\pi}{\sqrt{\pi}} \left[ \frac{\Gamma \left( \frac{n-1}{2}-\floor{\frac{n-3}{2}}\right)}{\Gamma \left( \frac{n}{2}-1+\lceil{\frac{3-n}{2}}\rceil \right)} + \frac{\Gamma \left( \frac{n}{2}+1-\floor{\frac{n}{2}}\right)}{\Gamma \left( \frac{n+1}{2} - \floor{\frac{n}{2}}\right)} - \frac{\Gamma \left( \frac{n+2}{2}\right)}{\Gamma \left(\frac{n+1}{2} \right)}-\frac{\Gamma \left( \frac{n+1}{2}\right)}{\Gamma \left(\frac{n}{2} \right)}\right]\Theta(n-1)  \nonumber \\
&~~~~~~~~~~~~~  + \sum _{m=1}^{n-1} \frac{\Gamma \left( \frac{n-m+1}{2} \right)}{\Gamma \left( \frac{n-m+2}{2}\right)} \left[ \frac{\Gamma \left( 2 + \floor{\frac{m-1}{2}}\right)}{\Gamma \left( \frac{3}{2} + \floor{\frac{m-1}{2}}\right)}+\frac{\Gamma \left( \frac{3}{2} + \floor{\frac{m}{2}}\right)}{\Gamma \left( 1 + \floor{\frac{m}{2}}\right)}\right]
\label{fun-sn}
\end{align}
Here 
$\floor{z}~(\text{or } \lceil z \rceil)$ denotes the greatest (or least) integer lesser (or greater) than or equal to $z$. {For large $n$, we find that $\mathcal{S}_n \simeq \pi n$ and inserting this in Eq. \eqref{fix-n-area} yields
\begin{align}
\langle A_n \rangle \simeq \frac{n \pi v_0^2}{2 \gamma ^2},~~~~\text{as } n \to \infty.
\label{fix-n-area-2}
\end{align}}

\item On the other hand, for fixed $t$ ensemble, we find that the mean area $\langle A(t) \rangle$ obeys the scaling relation
\begin{align}
\langle A(t) \rangle = \frac{v_0^2}{2 \gamma ^2} \mathcal{J}(\gamma t),
\label{fix-t-area}
\end{align}
where the scaling function $\mathcal{J}(z)$ is exactly given by
\begin{align}
\mathcal{J}(w) = e^{-w} \sum _{n=2}^{\infty}\frac{\mathcal{S}_n}{\Gamma (n+2)}w^{n+1}.
\label{fix-t-sca}
\end{align}

{The scaling function $\mathcal{J}(w)$ displays the following asymptotic behaviours:
\begin{align}
\mathcal{J}(w) & \simeq \frac{w^3}{3 \pi} +\frac{w^4}{4 \pi} \left(\frac{\pi}{6}-1 \right), ~~~~~~~\text{as } w \to 0, \\
& \simeq \pi w, ~~~~~~~~~~~~~~~~~~~~~~~~~~~~\text{as } w \to \infty.
\label{fix-t-eq-23-neww1}
\end{align}
Inserting these forms in Eq. \eqref{fix-t-area} yields that $\langle A(t) \rangle$ exhibits crossover from $\sim t^3$ scaling for $t \ll \gamma ^{-1}$ to $\sim t$ scaling for $t \gg \gamma ^{-1}$:
\begin{align}
\langle A(t) \rangle & \simeq \frac{\gamma v_0 ^2 t^3}{6 \pi},~~~~~\text{for }t\ll \gamma ^{-1} \label{fix-t-eq-24-neww1}\\
& \simeq \frac{\pi v_0^2}{2 \gamma} t,~~~~~~\text{for }t\gg \gamma ^{-1} .\label{fix-t-eq-25-neww1}
\end{align}}

The comparisons of the analytic expressions in  Eqs. \eqref{fix-n-area} and \eqref{fix-t-area} with numerical simulation are illustrated, respectively, in Figures \ref{fig-mean-area-1} and \ref{fig-mean-area-2}. 

\item We have also studied the variance and the distribution of the area in both ensembles numerically. We found that for large time ($n$ in fixed tumble ensemble and $t$ in fixed time ensemble) the variance of the area grows quadratically with time. In addition we found that the central part of the distribution (describing the typical fluctuations around the mean) for large time possesses a scaling form when the area is scaled by its mean and the scaling form matches with that of the Brownian motion, as expected. 
\end{enumerate}

\noindent
In what follows, we derive the results for the mean area explicitly and study it's typical fluctuations numerically.

\section{Mean area of the convex hull}
\label{mean-area-exact}
Let us begin by briefly summarising the central idea to compute the mean area of the convex hull for two dimensional stochastic processes. A more detailed account of this idea is given in {\cite{Furling2009, MajumdarFurling2010}}. Based on the knowledge of the Cauchy's formulae for closed curve {\cite{Cauchy1832}}, it was shown that the mean area and mean perimeter for random convex hull are related to the subject of extreme value statistics. To see this connection, consider a closed curve $\mathcal{C}$ parametrised by the points $\{(\mathcal{X}(s), \mathcal{Y}(s)) \}$ on its boundary where $s$ is the arc length. For the curve $\mathcal{C}$, we now define the support function $\mathcal{M}(\theta)$ along the direction $\theta$ (with respect to $x$-axis) as
\begin{align}
\mathcal{M}(\theta) = \underset{s \in \mathcal{C}}{\text{max}} \left[ \mathcal{X}(s) \cos \theta + \mathcal{Y}(s) \sin \theta \right].
\label{support-eq}
\end{align}
Geometrically, the support function $\mathcal{M}(\theta) $ represents the maximum extension of the curve $\mathcal{C}$ along the direction $\theta$. Interestingly, the perimeter and the area of the domain enclosed by $\mathcal{C}$ are given in terms of $\mathcal{M}(\theta)$ by Cauchy's formula as
\begin{align}
L &= \int _0 ^{2 \pi} d \theta~M(\theta), \label{perimeter} \\
A &= \frac{1}{2} \int _0 ^{2 \pi} d \theta \left[ \mathcal{M}^2 (\theta) - \left( \mathcal{M}'(\theta) \right)^2 \right].
\label{cauchy}
\end{align}
{To elaborate further, let us, for simplicity, consider a discrete time stochastic process of $n$ steps. Let} the positions of the particle at successive {(discrete)} times of a realisation {are denoted by} $\{ (X_i, Y_i) \}$ where $i=1,2,...,n$. We further consider that $\mathcal{C}$ now represents the convex hull corresponding to the points $\{ (X_i, Y_i) \}$. To construct the support function $\mathcal{M}(\theta)$ for $\mathcal{C}$, one clearly needs $\{(\mathcal{X}(s), \mathcal{Y}(s)) \}$ which is a difficult task. However, it was shown in {\cite{Furling2009, MajumdarFurling2010}} that this problem can be circumvented by noting the fact that $\mathcal{M}(\theta)$ is also the maximum of the projections of all points $\{ (X_i, Y_i) \}$ along the direction $\theta$. One can now write the support function $\mathcal{M}(\theta)$ as
\begin{align}
\mathcal{M}(\theta) &= \underset{1 \leq i \leq n}{\text{max}} \left[ X_i \cos \theta + Y _i \sin \theta \right]. 
\label{support-eq-2}
\end{align}
Using this form for $M(\theta)$ in Eqs.~\eqref{perimeter} and \eqref{cauchy}and then taking average over different realisations one gets the mean perimeter and the mean area of the convex hull $\mathcal{C}$. Since, we are interested in area only, we provide below the expression of mean area which follows directly from Eq. \eqref{cauchy}:
\begin{align}
\langle A _n \rangle = \frac{1}{2} \int _0 ^{2 \pi} d \theta \left[ \langle \mathcal{M}^2 (\theta) \rangle - \langle \left( \mathcal{M}'(\theta) \right)^2 \rangle \right].
\label{cauchy-eq-21}
\end{align} 
To proceed further, we assume that the maximum in Eq. \eqref{support-eq-2} is attained in the $k^*$-th step which enables us to write $\mathcal{M}(\theta) $ and $\mathcal{M}'(\theta)$ as
\begin{align}
&\mathcal{M}(\theta) = X _{k ^*} \cos \theta + Y _{ k ^*} \sin \theta,\label{support-eq-3}\\
& \mathcal{M}'(\theta) =- X _{k ^*} \sin \theta + Y _{ k ^*} \cos \theta.
\label{support-eq-4}
\end{align} 
For isotropic processes, the suppport function $\langle \mathcal{M}^2(\theta) \rangle$ and $\langle \mathcal{M}'^2(\theta) \rangle$ are independent of $\theta$ and we can consider just the direction $\theta = 0$. For this case, the mean area in Eq. \eqref{cauchy-eq-21} becomes 
\begin{align}
\langle A _n \rangle = \pi \left[ \langle M_n ^2 \rangle - {\langle Y_{k^*}^2 \rangle(n)  }\right],
\label{cauchy-eq-2}
\end{align} 
where $M _ n = \text{max} \left[ {X_1, X_2, ..., X_n} \right]$ is the maximum displacement along the $x$-axis and {$Y_{k^*}$} is the abscissa at $k^*$-th time-step at which the maximum $M _n$ along $x$-direction is reached. Later, Eq. \eqref{cauchy-eq-2} will be useful in calculating the mean area of the convex hull for RTP in fixed $n$ ensemble. 

Although Eq. \eqref{cauchy-eq-2} is derived for discrete time {isotropic stochastic} process, one can derive an analogous formula for the mean area of $\mathcal{C}$ in the continuous time case {\cite{MajumdarFurling2010}}. For this case, the mean area of the convex hull reads
\begin{align}
\langle A (t) \rangle = \pi \left[ \langle M^2(t) \rangle - {\langle Y(t _m)^2 \rangle(t)}  \right],
\label{cauchy-eq-3}
\end{align}
where $M(t)$ is the maximum of the $x$-coordinate till observation time $t$ i.e. $M(t) = \text{max}[\{X(\tau)\},~ \forall 0 \leq \tau \leq t]$ and $t_m$ is the time at which maximum $M(t)$ is reached. Also, $Y(t_m)$ {represents the $y$-coordinate of the RTP at time $t_m$ in trajectory of duration $t$.} Once again, Eq. \eqref{cauchy-eq-3} will be useful in computing the mean area for RTP in fixed $t$ ensemble.

Before closing this section, we remark that the formulae of mean area in Eqs. \eqref{cauchy-eq-2} and \eqref{cauchy-eq-3} apply to general isotropic $2$-$d$ stochastic process. 
{In the following,} we use these formulae to compute the mean area of $\mathcal{C}$ for {the RTP} model {in two dimension.} We {first} compute the mean area for the fixed-$n$ ensemble and then {focus on} the fixed-$t$ ensemble.

\section{Mean area for fixed-$n$ ensemble}
\label{mean-area-fixed-n-cal}
Let us first look at the RTP in fixed-$n$ ensemble where the total number of runs $n$ is fixed but the total time $t$ varies for different realisations. As indicated by  Eq. \eqref{cauchy-eq-2}, we need the maximum $M_n$ of the $x$-coordinate  trajectory
and the corresponding abscissa $Y_{k^*}  (n)$ to compute the mean area $\langle A_n \rangle$. 
{Recall from Eqs.~\eqref{x-cord-eq-2} and \eqref{y-cord-eq-2}, the position coordinates of the RTP performs random walks with correlated 
increments (jumps) $(x_i, y_i)$ are given in Eqs. \eqref{x-cord-eq-1} and \eqref{y-cord-eq-1}. Also recall that we have chosen the initial position of the RTP to be $(X_0,Y_0)=(0,0)$.} Let us first compute the joint probability distribution $p(x_i, y_i, \tau _i)$ of the increments {$x_i$ and $ y_i$ along, respectively, the $x$ and $y$ directions, and time duration} $\tau _i$ for the $i$-th run. Since the RTP moves ballistically during the time $\tau _i$, we have $v_0 \tau _i = \sqrt{x_i^2+y_i^2}$. Also, the time $\tau _i$ is exponentially distributed $\rho(\tau _i) = \gamma e^{-\gamma \tau _i}$. This enables us to write the joint distribution $p(x, y, \tau)$ as
\begin{align}
p(x,y, \tau)  = \frac{\gamma e^{-\gamma \tau}}{\pi} ~\delta(v_0 ^2 \tau ^2 - x^2-y^2),
\label{fix-n-eq-3}
\end{align}
where the factor $1/\pi$ comes from the normalisation condition. 
Finally, integrating $p(x,y, \tau)$ over $\tau$, we get the joint distribution of the increments $x$ and $y$ as
\begin{align}
p(x,y)  = \frac{\gamma}{2 \pi v_0~\sqrt{x^2+y^2}} ~~\text{exp}\left( -\frac{\gamma}{v} \sqrt{x^2+y^2}\right).
\label{fix-n-eq-4}
\end{align}
Notice that,  the problem of run and tumble motion {now got mapped} to a model of random walks in two dimensions with the jump distribution $p(x,y)$ given in Eq. \eqref{fix-n-eq-4}. Such mapping have  been considered in {\cite{Mori2019}} to study the persistent properties of RTP. The advantage now is that one can use Eq. \eqref{cauchy-eq-2}, true for discrete time processes, also for RTP. In what follows, we use this equation to compute the mean area of the convex hull for fixed-$n$ ensemble. From Eq. \eqref{cauchy-eq-2}, we see that this reduces to the problem of computing $\langle M_n^2 \rangle$ and {$\langle Y_{k^*}^2 \rangle(n)$} which we calculate below.

\subsection{Computation of $\langle M_n^2 \rangle$} 
{In order to compute the second moment of the maximum 
$M_n = \text{max} \{X_0, X_1,X_2,...,X_n \}$ 
of the $x$-component of a given trajectory of $n$ steps, we first recall that $\{X_0,X_1,X_2,...,X_n \}$} denote just a one dimensional random walk trajectory such that $X_i = X_{i-1}+x_i$. The increment $x_i$ is distributed according to the probability distribution $p_1(x_i)$ which is obtained by integrating the joint distribution $p(x_i,y_i)$ in Eq. \eqref{fix-n-eq-4} over all $y_i$. The resulting expression reads
\begin{align}
p_1(x) = \int _{- \infty}^{\infty} dy ~p(x,y) = \frac{\gamma}{\pi v_0} K_0 \left( \frac{\gamma |x|}{v_0}\right),
\label{fix-n-eq-5}
\end{align}
where $K_{\nu}(z)$ is the modified Bessel function of second kind. Note that $p_1(x)$ is both symmetric and continuous. Hence the random walker is characterised by the identical and independent increments $\{ x_i \}$ drawn from symmetric and continuous distribution $p_1(x_i)$.

To calculate $\langle M_n^2 \rangle$, we use the Pollaczek-Spitzer formula {\cite{Pollaczek1952, Spitzer1956}} which characterises the maximum $M_n$ for a random walk with identical and independent increments drawn from symmetric and continuous distribution. If $Q_n(M) = \text{Prob}[M_n \leq M]$ denotes the cumulative {probability} of $M_n$, then according to the Pollaczek-Spitzer formula, $Q_n(M)$ satisfies  {\cite{Grebenkov2017,ComtetA2005}}:
\begin{align}
\sum _{n=0}^{\infty} z^n \langle e^{-\lambda M_n} \rangle = \sum _{n=0}^{\infty} z^n \int_{0}^{\infty}d M e^{-\lambda M} Q_n'(M) = \frac{\phi (z, \lambda)}{\sqrt{1-z}},
\label{var-eq-1}
\end{align}
where $0 \leq z \leq 1$ and $\lambda \geq 0$ and the function $\phi(z, \lambda)$ is defined as
\begin{align}
&\phi(z, \lambda) = \text{exp}\left(-\frac{\lambda}{\pi} \int_{0}^{\infty} d \xi ~\frac{\ln(1-z \hat{p}_1(\xi))}{\lambda ^2 +k^2} \right), \label{var-eq-2}
\end{align}
{with $\hat{p}(\xi)$ being the Fourier transform of $p(x)$ given by} 
\begin{align}
&\hat{p}_1(\xi) = \int _{-\infty}^{\infty}d x ~e^{i \xi x} p_1(x) ~= \frac{1}{\sqrt{1+\xi^2 \sigma ^2}}.
\label{var-eq-3}
\end{align}
We have inserted $p_1(x)$ from Eq. \eqref{fix-n-eq-5} in writing $\hat{p}_1(\xi)$ and defined $\sigma = \frac{v_0}{ \gamma}$. One can suitably use Eq. \eqref{var-eq-1} to compute all moments of $M_n$. In fact, the Pollaczek-Spitzer formula was used in {\cite{Majumdar2010}} to determine the generating functions for all moments of $M_n$. For the first two moments, {one can show that}
\begin{align}
&h^{(1)}(z) = \sum_{n=0}^{\infty} z^n \langle M_n\rangle = \frac{1}{\pi (1-z)} \int_{0}^{\infty} \frac{d \xi}{\xi^2} ~\ln\left(\frac{1-z \hat{p}_1(\xi)}{1-z}\right), \label{var-eq-6} \\
&h^{(2)}(z) = \sum_{n=0}^{\infty} z^n \langle M_n^2\rangle = (1-z) \left[h^{(1)}(z) \right]^2 +\frac{\sigma ^2 z}{2(1-z)^2}.
\label{var-eq-7}
\end{align}
By appropriately differentiating $h^{(2)}(z)$ with respect to $z$, it is straightforward to show that the second moment $\langle M_n^2 \rangle$ {can be expressed completely in terms of the first moment $\langle M_n \rangle$ as}
\begin{align}
\langle M_n^2 \rangle = \sum _{m=1}^{n-1} \langle M_m \rangle \left[ \langle M_{n-m} \rangle-\langle M_{n-m-1} \rangle\right] + \frac{n \sigma ^2}{2}.
\label{fix-n-eq-6}
\end{align}
Expanding the right hand side of Eq.~\eqref{var-eq-6} one can in principle compute $\langle M_n \rangle$. Using Kac's formula {\cite{Kac1954}} for mean maximum displacement, it was recently computed explicitly in {\cite{Hartmann2020}} where it was shown that 
\begin{align}
\langle  M_n \rangle = \frac{\sigma }{2\sqrt{\pi}} \sum _{j=1}^{n} \frac{\Gamma\left( \frac{j+1}{2}\right)}{\Gamma\left( \frac{j+2}{2}\right)}.
\label{fix-n-eq-7}
\end{align} 
Using this expression, we get
\begin{align}
\langle M_{n-m} \rangle-\langle M_{n-m-1} \rangle = \frac{\sigma }{2\sqrt{\pi}} ~\frac{\Gamma\left( \frac{n-m+1}{2}\right)}{\Gamma\left( \frac{n-m+2}{2}\right)},~~n>1.
\label{fix-n-eq-8}
\end{align}
Finally inserting Eqs. \eqref{fix-n-eq-7} and \eqref{fix-n-eq-8} in the expression of $\langle M_n^2 \rangle $ in Eq. \eqref{fix-n-eq-6}  we get
\begin{align}
\langle M_n^2 \rangle = \frac{v_0 ^2}{2 \gamma ^2} \left( \frac{\mathcal{S}_n}{\pi}+n \right),~~n>1
\label{fix-n-eq-91}
\end{align}
with $\sigma = \frac{v_0}{\gamma}$ and 
\begin{align}
\mathcal{S}_n &= \frac{\sqrt{\pi}}{\sigma } \sum _{m=1}^{n-1} \frac{\Gamma\left( \frac{n-m+1}{2}\right)}{\Gamma\left( \frac{n-m+2}{2}\right)} \langle M _m \rangle. \label{fun-sn-22}
\end{align}
Inserting $\langle M _m \rangle$ from Eq. \eqref{fix-n-eq-7} in the above equation and simplifying, one gets the explicit expressions of $\mathcal{S}_n $ given in Eq.~\eqref{fun-sn}.

\subsection{Computation of $\langle  Y_{k^*}^2 \rangle(n)$}
\label{<Y_{k*}^2>}
{
We now compute the other term $\langle  Y_{k^*}^2 \rangle(n)$ in the expression of the mean area in Eq. \eqref{cauchy-eq-2}. To calculate this we first compute the joint distribution $\mathcal{P}(Y,k^*|n) = \text{Prob.}[Y_{k^*}=Y,k^*|n]$ and then compute the second moment of the displacement $Y_{k^*}$ of the particle along $y$ direction at step $k^*$ in which the particle reaches it's maximum along the $x$ direction in a walk of $n$-steps. It is possible to show that one can compute this joint distribution for a general $2$-d discrete time random walk  where the position coordinates $(X_i,Y_i)$ at  $i^{\rm th}$ step evolves, starting from $X_0=0,~Y_0=0$, as $X_i=X_{i-1}+x_i$ and $Y_i=Y_{i-1}+y_i$ with the jump increments $(x_i,y_i)$ at different steps are drawn independently from the common distribution $p(x,y)$. Note that at a given step, the increments $x$ and $y$ can be correlated. 
\\

\noindent
To proceed let us define marginal distribution of the $y$-increment 
\begin{align}
p_2(y) = \int_{-\infty}^\infty dx ~p(x,y), \label{p_2(y)}
\end{align}
and the marginal distribution of the $y$-coordinate $Y_k$ of the walker at step $k$
\begin{align}
P(Y,k) = \int_{-\infty}^\infty  \int_{-\infty}^\infty  ...\int_{-\infty}^\infty  &dy_1dy_2...dy_k ~\delta\left( Y- \sum_{i=1}^k y_i\right) ~\prod_{i=1}^k p_2(y_i). \label{P_marg(Y,k)}
\end{align}
Now consider any trajectory in $2$-d up to step $n$. Let $k^*$ denote the time at which the 
the $X_i$'s achieve their maximum $M_n$ and $Y_{k^*}$ denote the $y$-coordinate exactly
at step $k^*$. 
Recall, we want to compute the joint distribution of $Y_{k^*}$ and $k^*$, given the
total number of steps $n$, i.e., $\mathcal{P}(Y,k^*|n)$. We show that this joint distribution is given by 
\begin{equation}
\mathcal{P}(Y,k^*|n)= q_{k^*}\, q_{n-k^*} P(Y, k^*)\, ,
\label{mcalP(Y,k*)}
\end{equation}
where $P(Y, k^*)$ is defined in Eq.~\eqref{P_marg(Y,k)} and $q_n= {2n\choose n} \, 2^{-2n}$ is the Sparre-Andersen survival probability
of a $1$-d random walk with arbitrary symmetric and continuous jump distribution \cite{Andersen1954, Majumdar2010}.
To prove the claim in Eq.~\eqref{mcalP(Y,k*)}, we start with the joint probability distribution for $M_n=M,~Y_{k^*}=Y$ and $k^*$ denoted by $\mathscr{P}(M,Y,k^*|n)$. This joint probability  of the $n$-step walk can be expressed as a multi-dimensional integral
\begin{align}
& \mathscr{P}(M,Y,k^*|n) = \int \vec{dx}~\vec{dy}~\mathcal{Z} _{k^*}\big(M, Y, \{x_i \},\{y_i\} \big)\prod _{j=1}^{n} p(x_j, y_j), \label{fix-n-eq-10} 
\end{align}
with $\mathcal{Z} _{k^*}\left(M,Y, \{x_i \}, \{y_i\} \right)$ defined as
\begin{align}
\begin{split}
 \mathcal{Z} _{k^*}\big(M, Y, \{x_i \},\{y_i\} \big)=&~\Theta (M) \Theta (M-X_1) \Theta (M-X_2)...\Theta \left( M-X_{k^*-1}\right)  \times  \\
&\delta \left( M-X_{k^*}\right)~\delta \left( Y-Y_{k^*}\right) \Theta \left( M-X_{k^*+1} \right)...\Theta \left( M-X_{n} \right),
\end{split}
\label{fun-zk}
\end{align}
{where} $X_i = \sum _{j=1}^{i} x_j$ and $Y_i = \sum _{j=1}^{i} y_j$ {(recall we have chosen $X_0=Y_0=0$)}.  Here $\Theta(n)$ is the Heaviside theta function. 
The function $\mathcal{Z} _{k^*}\big(M, Y, \{x_i \},\{y_i\} \big)$ ensures that $X_{k^*}=M$ and $Y _{k^*} = Y$ while all other $\{X_i \}$ are smaller than $M$. 
Finally, we integrate over all $\{x_i,y_i\}$ with appropriate joint distribution $\prod_{i=1}^np(x_i, y_i)$. 
For simplicity, we have used the short-hand notation $\vec{dx}=dx_1 dx_2...dx_n$ and $\vec{dy}=dy_1 dy_2...dy_n$.

Since we are interested in the joint distribution $\mathcal{P}(Y,k^*|n) $ of $Y_{k^*}=Y$ and $k^*$, we integrate $\mathscr{P}(M,Y,k^*|n)$ in Eq. \eqref{fix-n-eq-10} over $M$ {\it i.e.}
\begin{align}
\mathcal{P}(Y,k^*|n) = \int_0^\infty dM~\mathscr{P}(M,Y,k^*|n). \label{new-eqq-2}
\end{align}
 To proceed further we take the Fourier transformation with respect to $Y$ 
 \begin{align}
\bar{\mathcal{P}} \left( \xi, k^*|n\right) &= \int _{-\infty} ^{\infty} d Y e^{i \xi Y} \int _{0}^{\infty} dM ~\mathscr{P}(M, Y, k^*|n), \label{fix-n-eq-11} 
\end{align} 
 and perform some algebraic simplications in Eq. \eqref{fix-n-eq-10}. We relegate the details of the calculations to  \ref{FTP} and present only the final result here. The final expression reads
\begin{align}
\bar{\mathcal{P}} \left( \xi, k^*|n\right) 
&= q _{k^*} q_{n-k^*}~\left[ \hat{p}_2(\xi)\right]^{k^*},
\label{fix-n-eq-12-new}
\end{align} 
where $\hat{p}_2(\xi)$ in Eq. \eqref{fix-n-eq-12-new} represents the Fourier transformation of the marginal distribution $p_2(y)$ of $y$-increment [see Eq.~\eqref{p_2(y)}] and is defined by
\begin{align}
& \hat{p}_2(\xi) = \int _{-\infty} ^{\infty} dy~e^{i \xi y} p_2(y). \label{new-eqq-3}
\end{align}
The term $q_n={2n\choose n} \, 2^{-2n}$, as mentioned earlier, is the survival probability of a random walker in one dimension starting from the origin  and with jumps drawn independently from a symmetric and continuous distribution. Note that the term $\left[ \hat{p}_2(\xi)\right]^{k^*}$ in Eq.~\eqref{fix-n-eq-12-new} is actually the Fourier transform of the marginal distribution $P(Y,k^*)$ of the $y$-coordinate at step $k^*$ as can be easily seen from Eq.~\eqref{P_marg(Y,k)}. Hence, performing inverse Fourier transform on both sides of Eq.~\eqref{fix-n-eq-12-new} one arrives at the result in Eq.~\eqref{mcalP(Y,k*)}.
Notice that the expression of $\mathcal{P}(Y,k^*|n)$ in Eq. \eqref{mcalP(Y,k*)}  appears naturally in the following form: 
\begin{align}
\begin{split}
\mathcal{P}(Y,k^*|n)=\text{Prob.}[Y_{k^*} =Y] ~\times~\text{Prob.}[M_n~\text{occurs~at~step}~k^*~\text{in}~n\text{-step~walk}].
\end{split}
\end{align} 
This result is 
quite universal and holds true for any joint distribution $p(x,y)$ as long as it is symmetric and continuous in $x$.
This universality is a consequence of Sparre Anderson theorem  \cite{Andersen1954, Majumdar2010}. In addition if the joint distribution $p(x,y)$ is such that $\langle y^2\rangle$ is finite, one finds
\begin{align}
\langle Y_{k^*}^2\rangle(n)  &= \sum_{k^* =1}^n  \langle Y_{k^*}^2 \rangle ~\times~q _{k^*} q_{n-k^*}  \label{<Y_k*^2>_1}\\ 
&= .\sum_{k^* =1}^n \langle y^2\rangle~k^*~q _{k^*} q_{n-k^*} \label{<Y_k*^2>_2}\\
&=\langle y^2\rangle ~\frac{n}{2}. \label{<Y_k*^2>_2}
\end{align} 
where we have used $q_k=2^{-2k} \binom {2k}{k}$. Note that this result is also universal. }

Recall, in this paper, we are interested to compute $\langle Y_{k^*}^2\rangle(n)$ for RTP in which case the joint distribution $p(x,y)$ is given in Eq. \eqref{fix-n-eq-4}.  For this distribution one has 
\begin{align}
& p_2(y) = \frac{\gamma}{\pi v_0} K_0 \left( \frac{\gamma |y|}{v_0}\right), \label{new-eqq-1} 
\end{align}     
with $\langle y^2\rangle=\frac{v_0^2}{\gamma^2}$ which gives 
\begin{align}
\langle Y_{k^*}^2\rangle(n) = \frac{v_0^2}{2\gamma^2}n. \label{fix-n-eq-13}
\end{align}

\begin{figure}[t]
\includegraphics[scale=0.27]{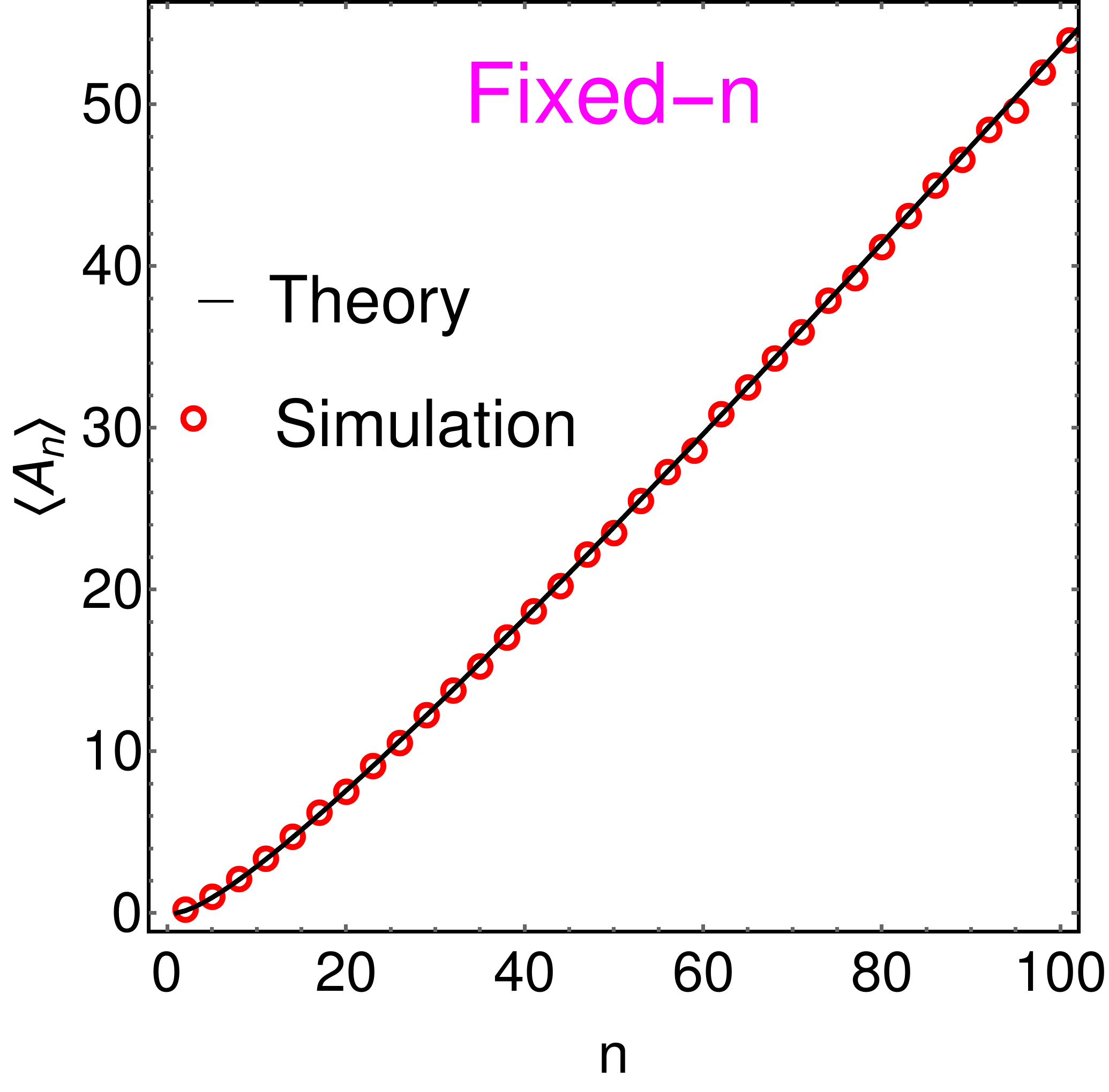}
\includegraphics[scale=0.28]{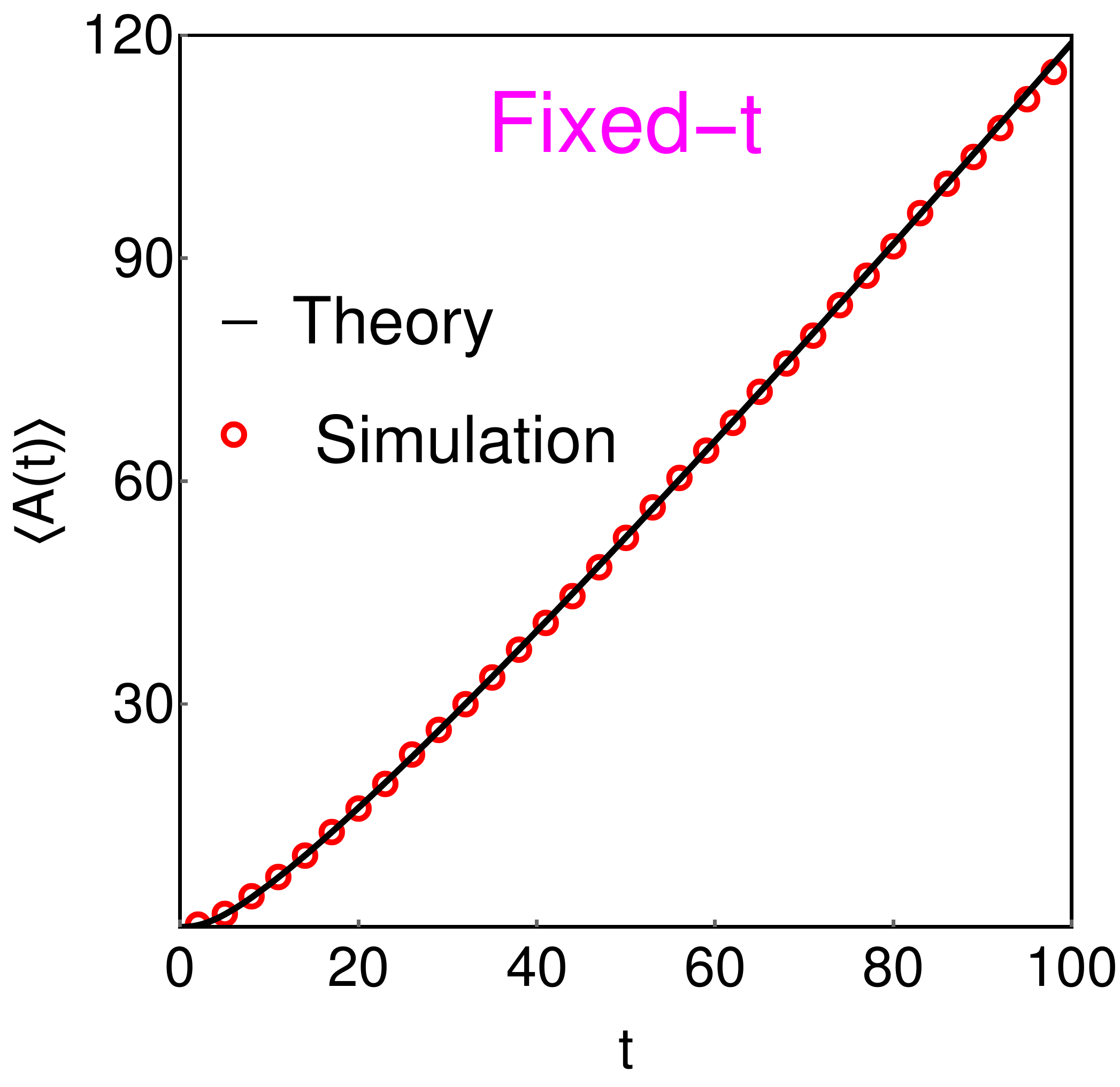}
\centering
\caption{Comparison of the mean area of the convex hull for a RTP in fixed-$n$ ensemble (left) and fixed-$t$ ensemble (right) with numerical simulations. The corresponding analytic expressions are given in Eqs. \eqref{fix-n-area} and \eqref{fix-t-area} respectively. Parameters chosen are $v_0=1$, $\gamma =1.5$ for left panel and $v_0=1$, $\gamma =1$ for right panel.}    
\label{fig-mean-area-1}
\end{figure}
\begin{figure}[t]
\includegraphics[scale=0.3]{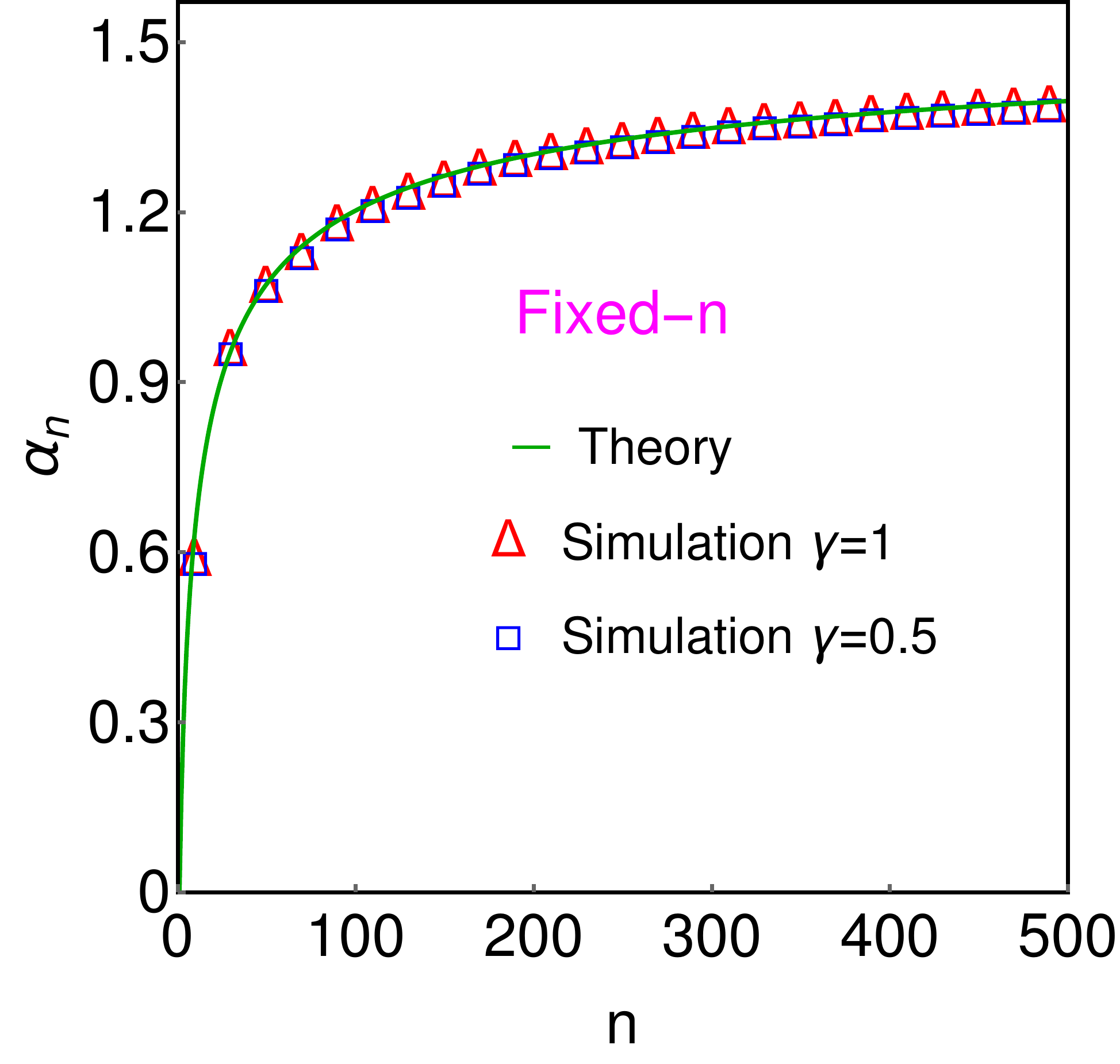}
\includegraphics[scale=0.3]{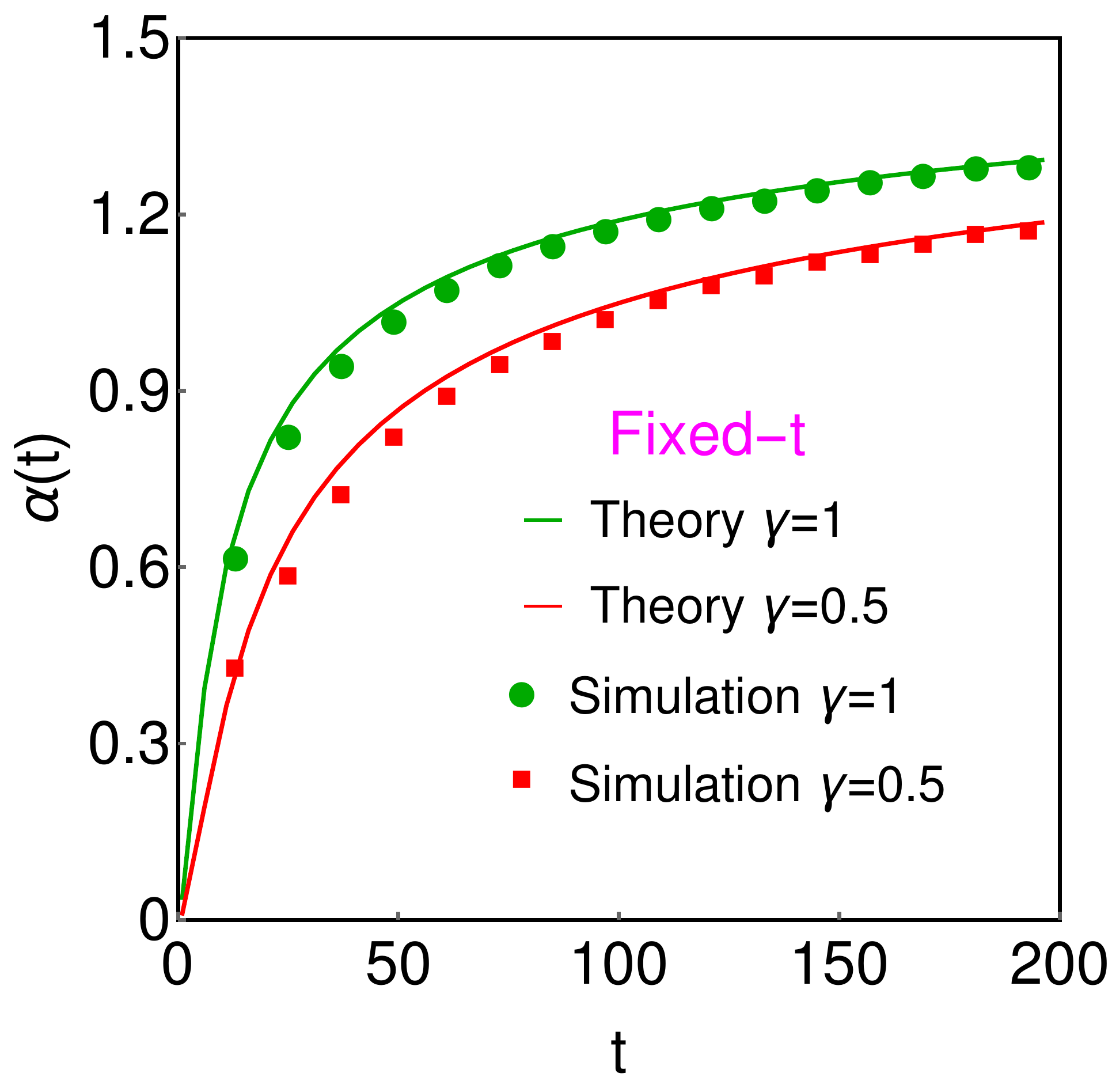}
\centering
\caption{Comparison of the mean area for fixed-$n$ (left) and fixed-$t$ (right) ensembles with the numerical simulations. For fixed-$n$ ensemble (left), we have denoted $\alpha _n = \langle A _n \rangle / n \sigma ^2$ with $\sigma = v_0 / \gamma $ and $\langle A_n \rangle $ given in Eq. \eqref{fix-n-area}. On the other hand, for fixed-$t$ ensemble (right), we have introduced the notation $\alpha (t) = \langle A(t) \rangle / \gamma t \sigma ^2$ with $ \langle A(t) \rangle$ given in Eq. \eqref{fix-t-area}. For both panels, we have used $v_0 =1$.}    
\label{fig-mean-area-2}
\end{figure}

\subsection{Mean area $\langle A_n \rangle$}
\label{mean-A_n}
Substituting {$\langle Y_{k^*} ^2 \rangle(n)$} from Eq. \eqref{fix-n-eq-13} along with $\langle M_n^2 \rangle $ from Eq. \eqref{fix-n-eq-91} in the expression of $\langle A_n \rangle$ in Eq. \eqref{cauchy-eq-2}, we obtain mean area of the convex hull for $2$-d isotropic run and tumble motion in fixed-$n$ ensemble as quoted in Eq. \eqref{fix-n-area}. For large $n$, we find $\mathcal{S}_{n} \simeq  \pi n$ (see \ref{largeSn} for proof) which yields the asymptotic form of $\langle A_n \rangle$ in Eq. \eqref{fix-n-area} as
\begin{align}
\langle A_n \rangle \simeq  \frac{n \pi}{2} \sigma ^2,~~~~~~~~\text{as } n \to \infty.
\label{fix-n-eq-131}
\end{align}
This matches with the mean area of the convex hull of a discrete two dimensional random walk of $n$ steps for any jump distribution with a finite variance $\sigma ^2$ {\cite{Grebenkov2017}}. In Figure \ref{fig-mean-area-1} (left panel), we have compared our analytic result of $\langle A_n \rangle$ in Eq. \eqref{fix-n-area} with the simulation results. We observe an excellent agreement between them. To construct convex hull numerically, we deploy the \textit{Andrew's monotone chain algorithm} {\cite{Andrew1979}} which is further expedited with Akl's heuristic {\cite{AKL1978}}. Then, to calculate the area, we denote the $m$ vertices of the convex hull as $\{ \bar{X}_i, \bar{Y}_i\},~1 \leq i \leq m$ in order of their Cartesian coordinates and use
\begin{align}
A = \frac{1}{2} \sum _{i=0}^{m -1} \left(\bar{Y}_i +\bar{Y}_{i+1} \right) \left(\bar{X}_i -\bar{X}_{i+1} \right),
\label{fix-n-eq-14}
\end{align}
with {$(\bar{X}_0, \bar{Y}_0) = (\bar{X}_{m},\bar{Y}_{m})$.} Finally, we estimate the mean area using the \textit{simple sampling} where we generate $10^4$ realisations of RTP, construct area of each of them using Eq. \eqref{fix-n-eq-14} and then take the average.

To compare the mean area for different parameters, we rescale $\langle A_n \rangle$ in Eq. \eqref{fix-n-area} with $\sigma ^2$ where $\sigma = v_0 / \gamma$. Moreover, from Eq. \eqref{fix-n-eq-131}, we see that $\langle A_n \rangle$ scales linearly with $n$ for $n \gg 1$. Therefore, we also rescale $\langle A_n \rangle$ with $n$ to remove the asymptotic growth with respect to $n$ for proper visualisation. Defining
\begin{align}
\alpha _n = \frac{\langle A_n \rangle}{ n \sigma ^2},
\label{new-fix-n-eq-15}
\end{align} 
one expects $\alpha _n$ to be independent of $v_0$ and $\gamma$ via Eq. \eqref{fix-n-area}. Also, $\alpha _n $ should saturate to the value $\frac{\pi}{2} \left( = 1.571.. \right)$ for $n \to \infty$. Indeed, in figure \ref{fig-mean-area-2} (left panel), we observe that $\alpha _n $ is identical for two different values of $\gamma $, namely $\gamma =1$ and $\gamma =0.5$. Moreoever, it approaches the value  $\frac{\pi}{2} $ as we go to higher values of $n$. This comparison of $\alpha _n$ for two different values of $\gamma$ provides another verification of $\langle A_n \rangle$ in Eq. \eqref{fix-n-area}.

\section{Fixed-$t$ ensemble}
\label{mean-area-fixed-t-cal}
The previous section dealt with the mean area of the convex hull for RTP in fixed-$n$ ensemble. We now consider the mean area in fixed-$t$ ensemble where the observation time $t$ is fixed but the number of runs $n$ varies from sample to sample. For this case also, we show that the run and tumble model can be suitably mapped to a random walker in two dimensions which is then used to calculate exactly the mean area $\langle A (t) \rangle$ via Eq. \eqref{cauchy-eq-3}. To begin with, let us consider a realisation of RTP with $n$ runs where the $i$-th run lasts for time $\tau _i$ with position increments $x_i$ and  $y_i$. Since at the end of each run except the $n$-th one, the RTP {encounters} a tumbling, the times $\{ \tau _i\}~\text{for}~1 \leq i \leq (n-1)$ are all drawn independently from the exponential distribution $\rho(\tau _i) = \gamma e^{-\gamma \tau _i}$. Therefore the joint distribution $p(x_i, y_i, \tau _i)$  with $1 \leq i \leq (n-1)$ is given by Eq. \eqref{fix-n-eq-3}. On the other hand, during the last interval $\tau _n$, the RTP does not encounter any tumble: the probability of which is $e^{-\gamma \tau _n}$. Hence, the corresponidng joint distribution is
\begin{align}
p_{\text{last}}(x_n,y_n, \tau _n)  = \frac{ e^{-\gamma \tau _n}}{\pi} ~\delta(v_0 ^2 \tau_n ^2 - x_n ^2-y_ n^2) = \frac{1}{\gamma} ~p(x_n, y_n, \tau _n),
\label{fix-t-eq-1}
\end{align} 
where $p(x_n, y_n, \tau _n)$ is given by Eq. \eqref{fix-n-eq-3}. We emphasise that unlike in the fixed-$n$ case, here the runs are correlated due to the constraint of fixed $t$. To see this more clearly, we write the grand joint distribution of $\{x_i\},~\{y_i\}$ and $n$ below:
\begin{align}
P\big(\{x_i\},\{y_i\}, n|t \big) = \int_0^td\tau_1 \int_0^t d\tau_2,...,\int_0^td\tau_n~\frac{1}{\gamma} \left[ \prod _{i=1}^{n} p(x_i, y_i, \tau _i) \right] \delta \left( \sum _{i=1}^{n} \tau _i - t\right).
\label{fix-t-eq-2}
\end{align}
To get rid of the $\delta$-function, we take Laplace transformation with respect to $t~(\to s)$ 
\begin{align}
\int _ 0^{\infty} dt ~e^{-st}~P\big(\{x_i\},\{y_i\}, n|t \big) &=\frac{1}{\gamma} \left[ \prod _{i=1}^{n}\frac{\gamma~\text{exp}\left( -\frac{(\gamma+s)}{v} \sqrt{x_i^2+y_i^2}\right)}{2 \pi v_0~\sqrt{x_i^2+y_i^2}} \right], 
\end{align}
which we rewrite as
\begin{align}
\int _ 0^{\infty} dt ~e^{-st}~P\big(\{x_i\},\{y_i\}, n|t \big)& = \frac{1}{\gamma} \left( \frac{\gamma}{\gamma +s}\right)^n~\left[ \prod _{i=1}^{n} g_{s}(x_i, y_i)\right], \label{fix-t-eq-3} \\
\text{with } ~~~~~~~g_s(x,y)& = \frac{(\gamma+s)~\text{exp}\left( -\frac{(\gamma+s)}{v} \sqrt{x_i^2+y_i^2}\right)}{2 \pi v_0 \sqrt{x^2+y^2}}.
\label{fix-t-eq-4}
\end{align}
Finally, inverting the Laplace transform in Eq. \eqref{fix-t-eq-3}, the grand joint distribution $P\big(\{x_i\},\{y_i\}, n|t \big)$ can be formally written as
\begin{align}
P\big(\{x_i\},\{y_i\}, n|t \big) = \int _{\Gamma} \frac{ds}{2 \pi i} ~e^{s t}~\frac{1}{\gamma} \left( \frac{\gamma}{\gamma +s}\right)^n~\left[ \prod _{i=1}^{n} g_{s}(x_i, y_i)\right],
\label{fix-t-eq-5}
\end{align} 
where $\Gamma$ is the Bromwich contour in the complex $s$ plane. 
{Note that the function $g_s(x,y)$ given in Eq.~\eqref{fix-t-eq-4} can be interpreted as a probability distribution as it is positive over full $(x,y)$ plane and normalised to unity. As a result the term inside the square bracket in the integrand of the Eq.~\eqref{fix-t-eq-5} can be interpreted as the joint distribution of the increments $x_i$ and $y_i$ of a random walker in two dimension in steps $i=1,2,...,n$. 
In the context of RTP such  mapping to random walk problem was observed  earlier \cite{Hartmann2020, MoriDoussal2020} and exploited to study the  survival probability  in higher dimension \cite{MoriDoussal2020}. In this paper we follow a similar calculation using this mapping and compute the mean area $\langle A (t) \rangle$ of the convex hull by employing the formula in Eq. \eqref{cauchy-eq-3}.} As seen in this formula, we need to calculate $\langle M^2(t) \rangle$ and {$\langle Y(t _m)^2 \rangle(t)$} to compute  $\langle A (t) \rangle$. In the following, we use the joint distribution $P\big(\{x_i\},\{y_i\}, n|t \big)$ in Eq. \eqref{fix-t-eq-5} to calculate these two quantities explicitly.

\subsection{Computation of $\langle M^2(t) \rangle$}
Let us begin with the computation of $\langle M^2(t) \rangle$ where $M(t)$ is the maximum of the $x$-co ordinate of RTP up to observation time $t$. For this, we need to compute the statistics of the maximum of a $1d$ random walker for fixed $t$ with $n$ jumps in the $x$-coordinates:  $\{ x_i\}$ for $i=1,2,...,n$. The joint distribution of increments $\{ x_i\}$ can be obtained by integrating $P\big(\{x_i\},\{y_i\}, n|t \big)$ in Eq. \eqref{fix-t-eq-5} over all $\{ y_i\}$ as
\begin{align}
P_x\big(\{x_i\}, n|t \big) &= \int _{-\infty} ^{\infty} dy_1 dy_2..dy_n ~P\big(\{x_i\},\{y_i\}, n|t \big), \\
& = \int _{\Gamma} \frac{ds}{2 \pi i} ~e^{s t}~\frac{1}{\gamma} \left( \frac{\gamma}{\gamma +s}\right)^n~\left[ \prod _{i=1}^{n} g_{s}(x_i)\right], \label{fix-t-eq-6} \\
\text{with } ~~~~~~~g_s(x)& = \int _{-\infty}^{\infty} dy~g_s(x,y) = \frac{(\gamma+s)}{\pi v_0} K_0 \left( \frac{(\gamma+s) |x|}{v_0}\right).\label{fix-t-eq-7}
\end{align} 
Using this expression of $P_x\big(\{x_i\}, n|t \big)$, we now proceed to calculate the statistics of the maximum $M(t)$. To this end, we define $Q(M,n|t)$ as the probability that $X_i <M$ for $1 \leq i \leq n$, where $X_i = \sum _{j=1}^{i} x_j$. It is  easy to realise that $Q(M,n|t)$ is actually the survival probability that the walker with $n$ steps up to time $t$ has not crossed $X=M$. Formally, this is given as

\begin{align}
Q(M,n|t) &= \int _{-\infty} ^{\infty} dx_1 ... \int _{-\infty} ^{\infty} dx_n~ \text{Prob.}\left[ X_1 <M, X_2 <M,...,X_n <M, n|t\right], \label{fix-t-eq-8} \\
& = \int _{-\infty} ^{\infty} dx_1 ... \int _{-\infty} ^{\infty} dx_n~\Theta (M-X_1)..\Theta (M-X_n)~P_x\big(\{x_i\}, n|t \big)
\label{fix-t-eq-9}
\end{align}
Note that $Q(M, n|t)$ is also the probability that the maximum displacement of the $1d$ random walk with $n$ steps up to time $t$ is $\leq M$. Differentiating $Q(M, n|t)$ with $M$ gives the joint probability distribution for $M$ and $n$ which can then be used to calculate $\langle M^2(t) \rangle$. The formal expression of $\langle M^2(t) \rangle$ reads 
\begin{align}
\langle M^2(t) \rangle &= \sum _{n=1}^{\infty} \int _{0}^{\infty} dM ~M^2~ \partial _M Q(M, n|t), \label{fix-t-eq-10} \\
& = \int _{\Gamma} \frac{ds}{2 \pi i} ~e^{s t}~\frac{1}{\gamma} \left( \frac{\gamma}{\gamma +s}\right)^n~ \langle M_s^2(n) \rangle
\label{fix-t-eq-11}
\end{align}
where $\langle M_s^2(n) \rangle$ is 
\begin{align}
\langle M_s^2(n) \rangle &= \int _{0}^{\infty} dM ~M^2~ \partial _M Q _s (M, n),~~~~~\text{with} \label{fix-t-eq-12} \\
Q _s (M, n) & = \int _{-\infty} ^{\infty} dx_1 ... \int _{-\infty} ^{\infty} dx_n~\Theta (M-X_1)..\Theta (M-X_n)~\left[ \prod _{i=1}^{n} g_{s}(x_i)\right].
\label{fix-t-eq-13}
\end{align}
Here $\int _{-\infty}^{\infty} dx~g_s(x)=1$ which can be verified easily from Eq. \eqref{fix-t-eq-7}. Hence $Q _s (M, n)$ can be deciphered as the cumulative distribution that the maximum is less than $M$ up to $n$ steps for an auxilliary $1d$ random walk with identical and independent jumps which follow symmetric and continuous distribution $g_s(x)$ given in Eq.~\eqref{fix-t-eq-7}. Consequently, $\langle M_s^2(n) \rangle$ is the second moment of the maximum $M_s(n)$ of the auxillary random walk which can be calculated using the Pollaczek Spitzer formulae in Eqs. \eqref{var-eq-6} and \eqref{var-eq-7} as done for fixed-$n$ ensemble. To avoid repetition, we present the details of this calculation in \ref{der-Ms2} and write only the final expression of $\langle M_s^2(n) \rangle$  here which reads
\begin{align}
\langle M_s^2(n) \rangle = \frac{v_0^2}{2(\gamma +s)^2} \left( \frac{\mathcal{S}_n}{\pi}+n\right),
\label{fix-t-eq-14}
\end{align}
where $\mathcal{S}_n$ is given in Eq. \eqref{fun-sn}. Substituting $\langle M_s^2(n) \rangle$ in the expression of $\langle M^2(t) \rangle$ in Eq. \eqref{fix-t-eq-11} and performing the inverse Laplace transformation gives
\begin{align}
\langle M^2(t) \rangle = \frac{v_0^2}{2 \gamma ^2} \left[ e^{-\gamma t}-1+ \gamma t +\frac{e^{-\gamma t}}{\pi} \sum _{n=1}^{\infty} \frac{\mathcal{S}_n}{\Gamma (n+2)} (\gamma t)^{n+1} \right].
\label{fix-t-eq-15}
\end{align}

\subsection{Computation of $\langle Y(t _m)^2 \rangle(t)$} 
\label{<Y(t _m)^2 >}
{
We next calculate $\langle Y(t _m)^2 \rangle(t)$ for the mean area $\langle A(t) \rangle$ in Eq. \eqref{cauchy-eq-3}. Recall that $Y(t_m|t)$ is the $y$-coordinate of the RTP at time $t_m$ when the maximum $M(t)$ of the $x$-coordinate is attained {in a trajectory of duration $t$.} To calculate $\langle Y(t _m)^2 \rangle(t)$, we first notice that 
for a trajectory of duration $t$ in the fixed-$t$ ensemble also the maximum in the $x$-direction occurs at the end of some complete jump step, say $k^*$ which is a function of the total number of jumps $n$ occurring in time $t$. Of course the number of jumps $n$ is a random quantity and consequently so is $k^*$ as they change from realisation to realisation and also they are functions of $t$. Hence denoting the time at the end of step $k^*$ by $t_m$, we can write $Y(t_m) = \sum _{i=1}^{k^*} y_i$.

We start with the  grand  joint distribution $P(\{x_i\},\{y_i\},n|t)$ given in Eq.~\eqref{fix-t-eq-5}. As we have mentioned earlier, the term $\prod_{i=1}^n g_s(x_i,y_i)$ inside the square bracket on the right hand side of this equation can be interpreted as the joint probability distribution of the jumps $x_i$ and $y_i$ for $i=1,2,...,n$ of a random walk in two dimension of $n$ steps. Once again we emphasise that $g_s(x,y)$, given explicitly in Eq.~\eqref{fix-t-eq-4}, can be interpreted as an effective joint distribution of elementary jumps along $x$ and $y$ directions, similar to $p(x,y)$ as considered earlier in sec.~\ref{mean-area-fixed-n-cal} except now it is parametrised by $s$. As a result we see that for a given trajectory of duration $t$ containing $n$ jump steps there is a trajectory of $n$ jumps generated by the joint distribution $\prod_{i=1}^n g_s(x_i,y_i)$. Hence, if the maximum displacement  in the $x$-direction occurs at step $k^*$ of a trajectory of duration $t$ containing $n$ jump steps, then in the auxiliary random walk problem generated by $g_s(x,y)$ the maximum displacement along $x$-direction occurs at the same step $k^*$.  Moreover the displacements $X_i$ and $Y_i$ (starting from the origin) along $x$ and $y$-directions at $i^{\rm th}$ step  are exactly same for $i=1,2,...,n$. Hence, we have 
\begin{equation}
\langle Y(t _m)^2 \rangle(t) = \int _{\Gamma} \frac{ds}{2 \pi i} ~e^{s t}~\frac{1}{\gamma}~\sum_{n=1}^\infty  \left( \frac{\gamma}{\gamma +s}\right)^n~\langle Y_{k^*}^2 \rangle_s(n), 
\label{<Y(t _m)^2>(t)}
\end{equation} 
where $\langle Y_{k^*}^2 \rangle_s(n)$ should be computed following the procedure given in sec.~\ref{<Y_{k*}^2>} with only difference being  the joint distribution $p(x,y)$ is replaced by $g_s(x,y)$ which is given in Eq.~\eqref{fix-t-eq-4}. That is why we now have a subscript $s$ in the notation of $\langle Y_{k^*}^2 \rangle_s(n)$.  Executing the computation steps from Eqs.~\eqref{<Y_k*^2>_1} - \eqref{<Y_k*^2>_2}  with $g_s(x,y)$ we get 
\begin{align}
\langle Y_{k^*}^2 \rangle_s(n) = \frac{v_0^2}{(\gamma+s)^2}\frac{n}{2}, 
\label{Y_k-s}
\end{align}
where we have used $\langle y^2 \rangle_{g_s}  =  \int _{-\infty}^{\infty}  dx  \int _{-\infty}^{\infty} dy~ y^2 ~ g_{s}(x, y) = \frac{v_0^2}{(\gamma +s)^2} $.
Inserting the above expression from Eq.~\eqref{Y_k-s} in Eq.~\eqref{<Y(t _m)^2>(t)} and carrying out the sum over $n$ we get 
\begin{align}
\langle Y(t _m)^2 \rangle(t) 
&=\frac{v_0^2}{2}\int _{\Gamma} \frac{ds}{2 \pi i} ~ \frac{e^{s t}}{ s^2 (\gamma +s)}.
\end{align}
which upon performing inverse Laplace transformation with respect to $s$ gives the final expression
\begin{align}
\langle Y(t _m)^2 \rangle(t) 
& =\frac{v_0^2}{2 \gamma ^2} \left( \gamma t-1+e^{-\gamma t}\right).
\label{fix-t-eq-23-n}
\end{align}
}

\subsection{Mean area for fixed-$t$ ensemble} 
The expressions of $\langle M^2(t) \rangle$ and {$\langle Y(t _m)^2 \rangle(t) $} in Eqs. \eqref{fix-t-eq-15} and \eqref{fix-t-eq-23-n} respectively guide us to write the mean area $\langle A(t) \rangle$ via Eq. \eqref{cauchy-eq-3}. Inserting these forms explicitly, it is straightforward to show that $\langle A(t) \rangle$ indeed possesses the scaling form of Eq. \eqref{fix-t-area} with the scaling function $\mathcal{J}(w)$ given in Eq. \eqref{fix-t-sca}. In Figure \ref{fig-mean-area-1} (right panel), we have plotted $\langle A(t) \rangle$ and compared against the numerical simulations. We observe excellent agreement. Here also, the convex hull is constructed numerically by using \textit{Andrew's monotone chain algorithm} and we have used Eq. \eqref{fix-n-eq-14} to calculate the area.

To get the Brownian limit of the expression of $\langle A(t) \rangle$ in Eq. \eqref{fix-t-area}, we look at the asymptotic behaviours of the scaling function $\mathcal{J}(w)$ which read 
\begin{align}
\mathcal{J}(w) & \simeq \frac{w^3}{3 \pi} +\frac{w^4}{4 \pi} \left(\frac{\pi}{6}-1 \right), ~~~~~~~\text{as } w \to 0, \\
& \simeq \pi w, ~~~~~~~~~~~~~~~~~~~~~~~~~~~~\text{as } w \to \infty.
\label{fix-t-eq-23}
\end{align}
Inserting these forms in Eq. \eqref{fix-t-area}, we find that $\langle A(t) \rangle$ exhibits crossover from $\sim t^3$ scaling for $t \ll \gamma ^{-1}$ to $\sim t$ scaling for $t \gg \gamma ^{-1}$:
\begin{align}
\langle A(t) \rangle & \simeq \frac{\gamma v_0 ^2 t^3}{6 \pi},~~~~~\text{for }t\ll \gamma ^{-1} \label{fix-t-eq-24}\\
& \simeq \frac{\pi v_0^2}{2 \gamma} t,~~~~~~\text{for }t\gg \gamma ^{-1} .\label{fix-t-eq-25}
\end{align}
For $t \gg \gamma ^{-1}$, we recover the result for Brownian motion with effective diffusion constant $D = v_0^2 /2 \gamma$. However, at small times, the behaviour is remarkably different than that of the Brownian motion as illustrated by the $\sim t^3$ growth in Eq. \eqref{fix-t-eq-24}. 
 This qubic growth can be easily understood by noting that at small times the RTP experiences only few tumbles. The minimum number of tumbling required to constract a convex hull is two tumbling events till time $t$ (counting the starting point as a tumble). Then, the convex hull is essentially a triangle with two sides of length $v_0 \tau $ and $v_0 (t- \tau)$ and some angle $\zeta$ between them. The area is given by $A(t) =| \frac{ v_0^2 \tau (t-\tau) \sin \zeta}{2}| $. To calculate mean, we recall that $\tau$ is drawn from exponential distribution $p(\tau) = \gamma e^{-\gamma \tau}$ and $\zeta$ is chosen uniformly from $[0, 2 \pi]$. It is then easy to show that the resulting mean exactly matches with the short time asymptotics in Eq. \eqref{fix-t-eq-24}. Although at large times, $\langle A(t) \rangle$ behaves identical to that of the Brownian motion, the short time behaviour is rather different. Another way to demonstrate this difference is to define
\begin{align}
\alpha (t) = \frac{ \langle A(t) \rangle}{\gamma t \sigma ^2}.
\label{fix-t-eq-24-new}
\end{align}
For $t \to \infty$, $\alpha(t)$ saturates to the value $\frac{\pi}{2}$. In Figure \ref{fig-mean-area-2} (right panel), we have plotted $\alpha(t)$ for two different values of $\gamma$ and also compared them against the numerical simulations. We see agreement of the numerical data to the analytic expressions for both cases. Also, we obtain that $\alpha(t)$ approaches the value $\frac{\pi}{2} $ in both cases.

\begin{figure}[t]
\includegraphics[scale=0.35]{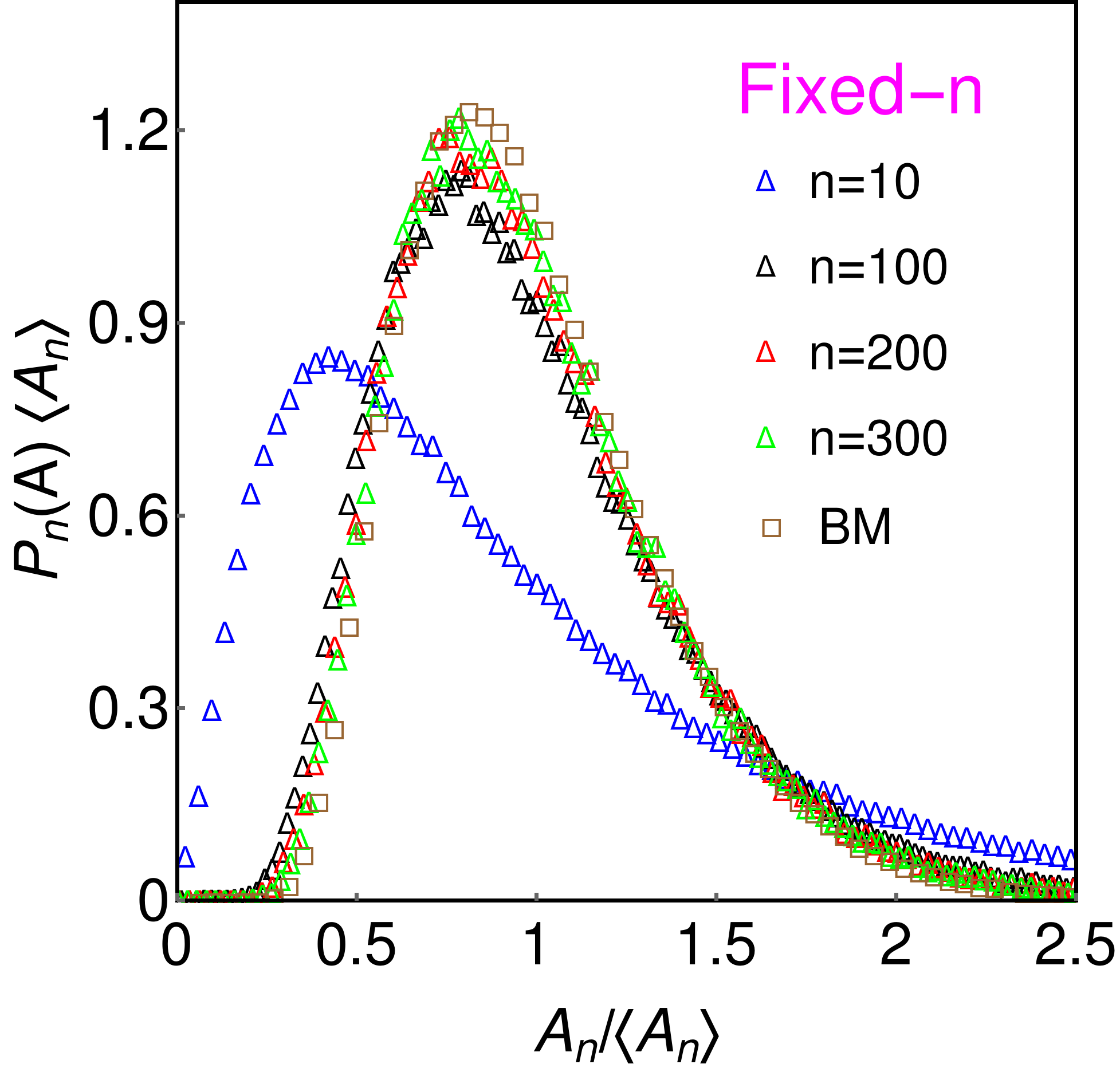}
\includegraphics[scale=0.35]{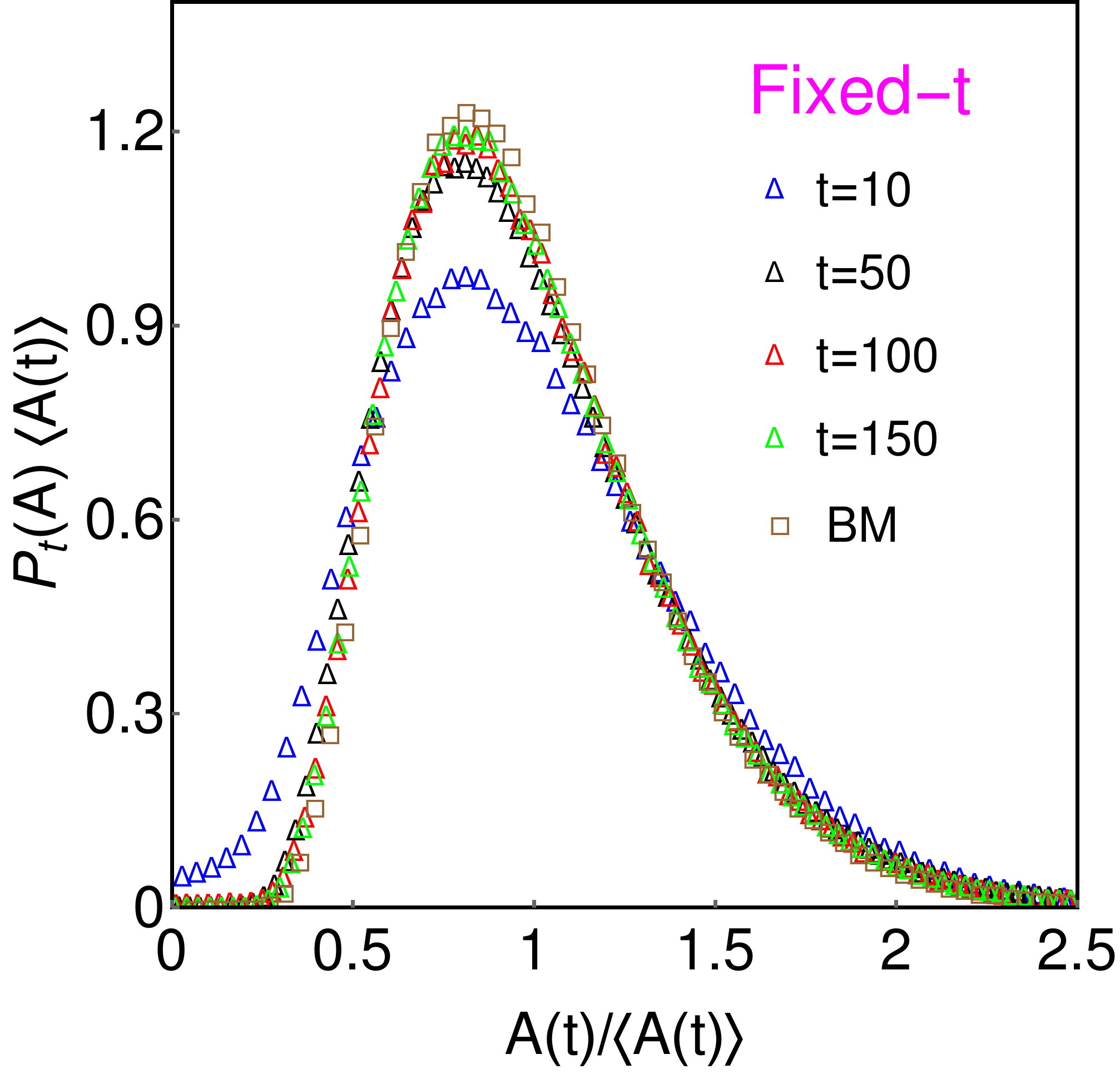}
\centering
\caption{Left: Simulation results for the distribution of the area $A_n$ for fixed-$n$ ensemble for different values of $n$. We have rescaled the distribution with mean area $\langle A_n \rangle$ and compared it with that of the Brownian motion. We have chosen $v_0=1$ and $\gamma =1$ for all values of $n$. Right: The same analysis is conducted for the fixed-$t$ ensemble with same choice of parameters.}    
\label{Fig-area-dist}
\end{figure} 
\section{Numerical study of the probability distribution}
\label{numerical dist}
In the previous sections, we explicitly derived the exact analytic expressions of the mean area of the convex hull in fixed-$n$ and fixed-$t$ ensembles and compared them against the numerical simulations. These expressions are given respectively in Eqs. \eqref{fix-n-area} and \eqref{fix-t-area}. We now investigate the probability distribution of the area of the convex hull for a single RTP. Deriving analytic forms of the distribution seems a difficult problem. In view of this, we perform a rigorous numerical study for the distribution in the two statistical ensembles. Here, we only look at the distribution corresponding to the typical fluctuations in area. By this, we mean the parts of distribution that lie within few standard deviations around the mean. To compare the distribution for different values of $n$ or $t$, it turns out useful to rescale it with the mean area. In Figure \ref{Fig-area-dist}, we have illustrated the simulation data for the rescaled distribution for different values of $n$ and $t$. For both ensembles, we find that the distribution converges to that of the Brownian motion in the asymptotic regime, i.e. $n \gg 1$ for fixed-$n$ ensemble and $t \gg \gamma ^{-1}$ for fixed-$t$ ensemble. However, for other (small and intermediate) values of $n$ and $t$, we expectedly see clear departure from the Brownian motion as elucidated by blue symbols in both panels of Figure \ref{Fig-area-dist}. To construct the distributions in Figure \ref{Fig-area-dist}, we have adopted the simple sampling techniques where we take a realisation of RTP depending on the ensemble that we are interested in. Given this trajectory, we construct the convex hull using \textit{Andrew's monotone chain algorithm} along with Akl's heuristic and use Eq. \eqref{fix-n-eq-14} to calculate the area. This procedure is then repeated for $10^{5}$ realisations to finally construct the histogram. 

\begin{figure}[t]
\includegraphics[scale=0.33]{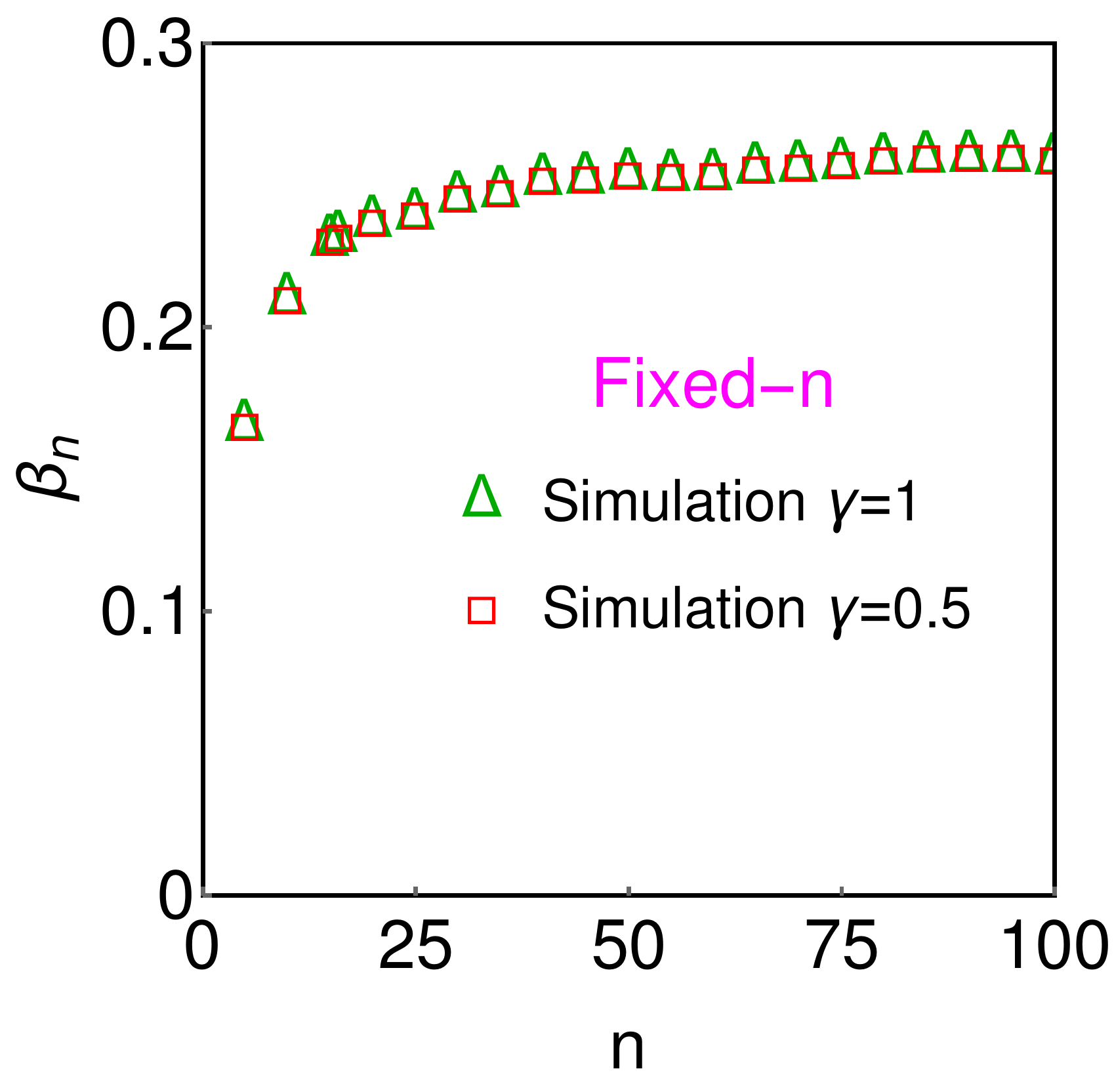}
\includegraphics[scale=0.3]{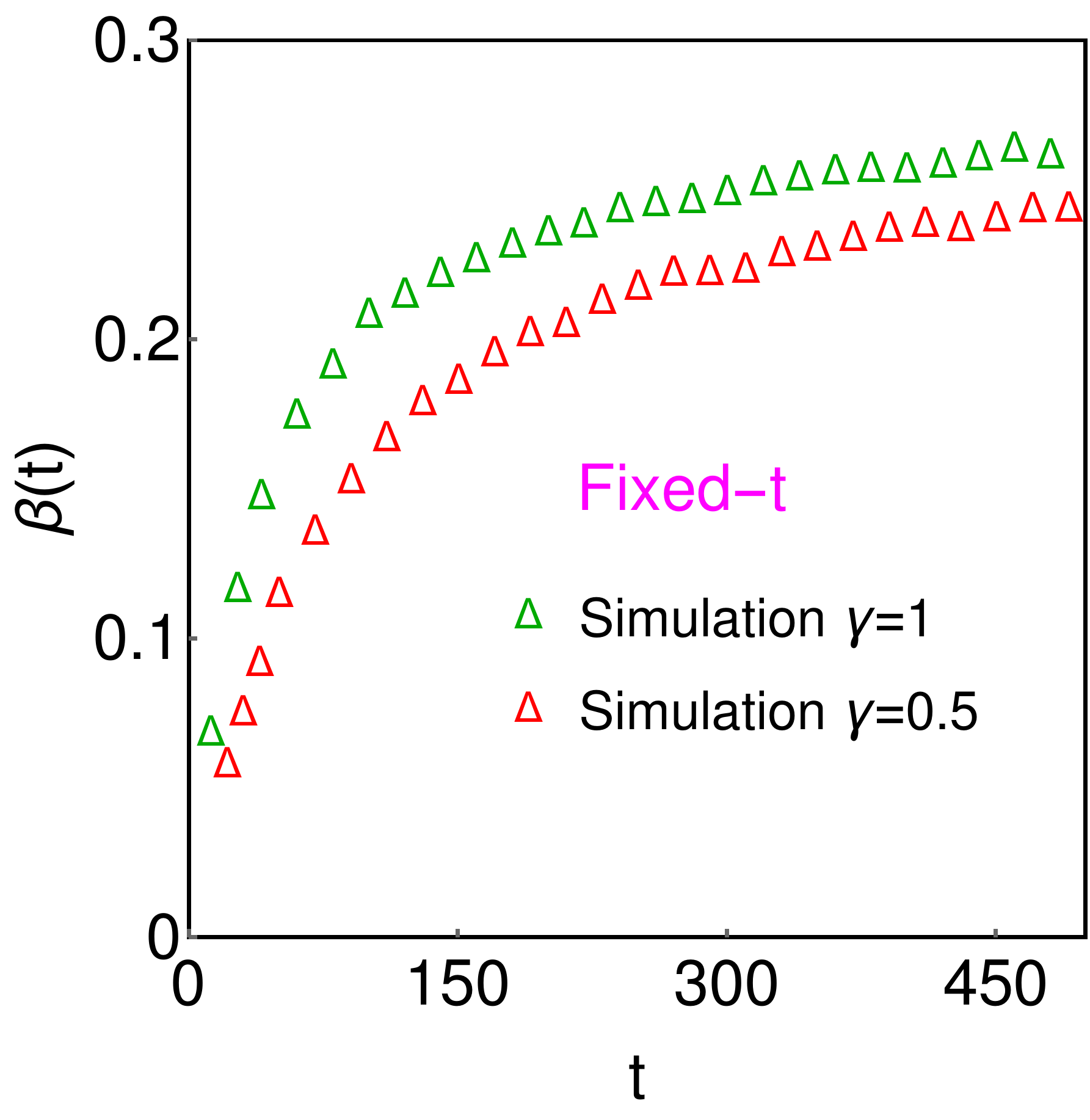}
\centering
\caption{Simulation data for the variance of area for fixed-$n$ (left) and fixed-$t$ (right) ensembles. For fixed-$n$ ensemble (left), we have plotted $\beta _n = \text{Var}(A_n)/n^2 \sigma ^4$ {\it vs.} $n$ and for fixed-$t$ ensemble (right), we have plotted $\beta (t) =\text{Var}(A(t))/(\gamma t)^2 \sigma ^4 $ {\it vs.} $t$. Parameters used in these plots are $v_0=1$, $\gamma =1$ (green) and $\gamma =0.5$ (red) for both panels.}    
\label{Fig-area-var}
\end{figure}

Similarly, we have also studied the variance of the area in Figure \ref{Fig-area-var} for two ensembles. As done for the mean area in Eqs. \eqref{new-fix-n-eq-15} and \eqref{fix-t-eq-24-new}, we define the following two quantities:
\begin{align}
\beta _n & = \frac{\text{Var}(A_n)}{n^2 \sigma ^4},~~~~~~~~~\text{for fixed-}n \\
\beta (t) &=\frac{\text{Var}(A(t))}{(\gamma t)^2 \sigma ^4}. ~~~~~~~~~\text{for fixed-}t
\end{align}
As seen before, this rescaling of the variance helps in better visualisation of the data since all of them converge to the same value in the asymptotic regime for both ensembles and for different values of the parameters. For both ensembles, we see in Figure \ref{Fig-area-var} that $\beta _n$ and $\beta (t)$ tend towards the same value for different values of $\gamma$.

\section{Conclusion}
\label{conclusion}
We have investigated the area of the convex hull of a run-and-tumble particle in two dimensions. We have considered this problem in two different ensembles: (i) fixed-$n$ ensemble and (ii) fixed-$t$ ensemble. We have obtained explicit expressions of the mean area $\langle A_n \rangle$ and $\langle A(t) \rangle$ in these two ensembles and verified them numerically. 
To study mean area analytically, we have used a mapping of the  run-and-tumble motion to a random walk model in two dimensions similar to what was used previously in \cite{MoriDoussal2020}. After exploiting the connection between the extreme value statistics and the computation of the mean area through Cauchy formulae [Eqs.~\eqref{cauchy-eq-2} and \eqref{cauchy-eq-3}], we use this mapping to employ the Sparre Anderson theorem which finally lead us arrive at the explicit expressions of the mean area in Eqs. \eqref{fix-n-area} to \eqref{fix-t-sca}. We observed that at large times the mean area grows linearly whereas at small times it grows as $\sim t^3$ with $t$. We have obtained a scaling function that describes the crossover from the qubic growth to linear growth around the natural time scale $\gamma^{-1}$ provided by the tumbling rate. 

Obtaining analytic results for higher order moments and distribution seems a challenging task. We have numerically studied the variance of the mean area as time ( number of tumbles in the fixed ensemble and $t$ in the fixed time ensemble) and found that it grows quadratically with time at long times. {We have also studied the distribution of the area numerically. While in the asymptotic regime, i.e. $n \gg 1$ for fixed-$n$ ensemble and $t \gg \gamma ^{-1}$ for fixed-$t$ ensemble, the distribution converges to that of the Brownian motion when area is scaled with its mean, we find clear difference at small or intermediate regime.}

{As mentioned before, computing the higher order moments and the full distribution of the area is a challenging problem and still remains an open problem even for Brownian particle.}
It is worth emphasizing that we have looked at the simple version of run and tumble model where tumbles are instantaneous. However, it has been experimentally found that active particles in reality spends small but non-zero time while tumbling {\cite{Basu2020, Berg2003}}. Extending our results for these realistic systems remains a promising future direction. Finally, in this work, we have focused on one run and tumble model active particles. It would be interesting  to explore how our results get generalised for other models of active particles like active Brownian particle and active Ornstein-Uhlenbeck particle {\cite{Romanczuk2012, Bechinger 2016}}.

\section{Acknowledgement}
{AK and PS acknowledge support of the Department of Atomic Energy, Government of India, under project no.12-R\&D-TFR-5.10-1100. AK acknowledges support from DST, Government of India grant under project No. ECR/2017/000634. We thank B. De Bruyne and F. Mori for useful discussions.
}

\appendix

\section{Derivation of $\bar{\mathcal{P}} \left( \xi, k^*|n\right) $ in Eq. \eqref{fix-n-eq-12-new}}
\label{FTP}
{
In this appendix we derive the expression of $\bar{\mathcal{P}} \left( \xi, k^*|n\right) $ given in Eq. \eqref{fix-n-eq-12-new}. To begin with, we start with the joint distribution $\mathscr{P}(M,Y,k^*|n)$ in Eq. \eqref{fix-n-eq-10}. Performing Fourier transform with respect to $Y$, one can write $\bar{\mathcal{P}} \left( \xi, k^*|n\right)$ defined in Eq.~\eqref{fix-n-eq-11} as
\begin{align}
\bar{\mathcal{P}} \left( \xi, k^*|n\right) &= \int_0^\infty dM~I_{\rm left}(M,\xi,k^*)~I_{\rm right}(M,k^*,n),  
\label{mcal_P_1}
\end{align} 
where we have defined
\begin{align}
I_{\rm right}(M,k^*,n) = \int \prod_{i=k^*+1}^N dy_i\, dx_i\, p(x_i,y_i)\, \Theta \left(M-\sum_{j=1}^i x_j \right), \label{I_right}
\end{align}
and 
\begin{align}
I_{\rm left}(M,\xi,k^*) = \int \prod_{i=1}^{k^*} dx_i\, {\tilde p}(x_i, \xi)\, \prod_{i=1}^{k^*-1} \Theta\left(M-\sum_{j=1}^i x_j\right)
\delta\left(M-\sum_{j=1}^{k^*} x_j\right)\, , \label{I_left}
\end{align}
with the definition
\begin{equation}
{\tilde p}(x,\xi)= \int_{-\infty}^{\infty} dy e^{i \xi y} p(x,y) \, .
\label{pxk.1}
\end{equation}  
Let us first consider the integral $I_{\rm right}(M, k^*,n)$ defined in Eq. (\ref{I_right}).
Since, 
\begin{equation}
X_{k^*+j}= x_1+x_2+\ldots+ x_{k^*}+ x_{k^*+1}+\ldots+ x_{k^*+j}= M + x_{k^*+1}+\ldots+ x_{k^*+j}
\label{Xr.1}
\end{equation}
upon using $M= \sum_{i=1}^{k^*} x_i$, we can re-write Eq. (\ref{I_right}) as
\begin{equation}
I_{\rm right}(M, k^*, n)=\int \left[\prod_{i=k^*+1}^n dx_i p_1(x_i)\right] 
\prod_{i=k^*+1}^n \Theta \left(- \sum_{j=k^*+1}^i x_{j}\right)\,
\label{Xr.2}
\end{equation}
where $p_1(x)= \int_{-\infty}^{\infty} p(x,y)\, dy$ is a normalized (to unity) probability density function for
the increment in the $x$-direction. However, the integral in Eq. (\ref{Xr.2}) is simply the probability
that a random walk in one dimension (in the $x$-direction) starting at the origin, with independent and identically distributed increment $x_i$'s drawn from $p_1(x_i)$,
stays below the origin up to step $n-k^*$. This is precisely given by $q_{n-k^*}$ via the Sparre Andersen theorem
(independently of the jump distribution $p_i(x)$, where $q_n= {2n \choose n}\, 2^{-2n}$. Hence, we have
\begin{equation}
I_{\rm right}(M, k^*, n)= q_{n-k^*} \, .  
\label{Ir.3}
\end{equation}
Note that the integral $I_{\rm right}(M, k^*, n)$ does not depend on $M$, but only on $(n-k^*)$.

We now turn to the left integral $I_{\rm left}(M,\xi,k^*)$ in Eq. (\ref{I_left}). 
Let us first re-write ${\tilde p}(x,\xi)$ in Eq. (\ref{pxk.1}) in a different way.
Let us first consider the integral
\begin{equation}
\int_{-\infty}^{\infty} dx \,{\tilde p}(x,\xi)=
\int_{-\infty}^{\infty} dx\, \int_{-\infty}^{\infty} dy\, p(x,y)\, e^{i \xi y}
=\int_{-\infty}^{\infty} dy\, p_2(y)\, e^{i \xi y}= {\tilde p}_2(\xi) \, ,
\label{rw.1}
\end{equation}
where $p_2(y)= \int_{-\infty}^\infty dx\, p(x,y)$ is the marginal distribution for the $y$-increment. Now, let us re-write
\begin{equation}
{\tilde p}(x,\xi)= \frac{ {\tilde p}(x,\xi)}{ \int_{-\infty}^{\infty} dx\, {\tilde p}(x,\xi)} 
\times {\tilde p}_2(\xi)= f(x,\xi)\, {\tilde p}_2(\xi)
\label{pxk.2}
\end{equation}
where we used the identity in Eq. (\ref{rw.1}) and
\begin{equation}
f(x,\xi)= \frac{{\tilde p}(x,\xi)}{\int_{-\infty}^{\infty} dx\, {\tilde p}(x,\xi)}\, .
\label{fxk.1}
\end{equation}
Note that $f(x,\xi)$ is normalized to unity (when integrated over  $x$) and can be thought of as an effective
jump distribution in the $x$ direction that is just parametrized by $\xi$ assuming it is positive for all $x$.
We use this expression of ${\tilde p}(x,\xi)$ from Eq. (\ref{pxk.2}) into the integral
expression for $I_{\rm left}(M,\xi,k^*)$ in Eq. (\ref{I_left}), to get
\begin{equation}
I_{\rm left}(M,\xi,k^*)= \left[{\tilde p}_2(\xi)\right]^{k^*}\,
\int \prod_{i=1}^{k^*} dx_i\, f(x_i, \xi)\, \prod_{i=1}^{k^*-1} \Theta\left(M-\sum_{j=1}^i x_j\right)
\delta\left(M-\sum_{j=1}^{k^*} x_j\right).
\label{Il.2}
\end{equation}
Now, substituting this expression and the result in Eq. (\ref{Ir.3}) on the right hand side
of Eq. (\ref{mcal_P_1}) and carrying out the integral over $M$ gives
\begin{align}
\begin{split}
{\tilde P}(\xi,k^*|n)= q_{n-k^*}\,& \left[{\tilde p}_2(\xi)\right]^{k^*}\,
\int \left[\prod_{i=1}^{k^*} dx_i\, f(x_i, \xi)\right]\, \prod_{i=1}^{k^*}\Theta\left(\sum_{j=1}^i x_{k^*+1-j} \right)
\end{split}
\label{Pf.1}
\end{align}
However, we immediately identify the $k^*$-fold integral in Eq. (\ref{Pf.1}) as the probability that
a one dimensional random walker, starting at the origin and with jump distribution drawn from
$f(x,\xi)$ (which is normalised to unity), stays above the origin up to $k^*$ steps. 
By Sparre Andersen theorem, this is universal and is simply $q_{k^*}$ and is independent
of $f(x,\xi)$, and in particular then does not depend on $\xi$. Hence we finally have
\begin{equation}
{\tilde P}(\xi, k^*|n)= \int_{\infty}^{\infty} P(Y, k^*|n)\, e^{i \xi Y}\, dY=
q_{k^*}\, q_{n-k^*}\, {\left[{\tilde p}_2(\xi)\right]}^{k^*} \, ,
\label{th.3}
\end{equation}
which upon Fourier inversion, yields the result in Eq. (\ref{mcalP(Y,k*)}).  This result is 
 true for arbitrary joint distribution $p(x,y)$ as long as it is symmetric and continuous in $x$.
 }

\section{Proof $\mathcal{S}_n \simeq \pi n$ as $n \to \infty$}
\label{largeSn}
In this appendix, we derive the asymptotic form of $\mathcal{S}_n$ for large $n$ which was used to obtain the large $n$ behaviour of $\langle A_n \rangle$ in Eq. \eqref{fix-n-eq-131}. To this end, we consider the expression of $\mathcal{S}_n$ in Eq. \eqref{fun-sn-22} and change the variable $m = z n$ to yield
\begin{align}
\mathcal{S}_n &= \frac{\sqrt{\pi}}{\sigma } \sum _{z=1/n}^{(n-1)/n} ~~\frac{\Gamma\left( \frac{n(1-z)+1}{2}\right)}{\Gamma\left( \frac{n(1-z)+2}{2}\right)} ~\langle M _{n z} \rangle.  \label{largeSn-eq-1}
\end{align}
Note that $z \in \{\frac{1}{n},\frac{2}{n},...,\frac{n-1}{n} \}$. For large $n$, we change the summation in Eq. \eqref{largeSn-eq-1} to integration as $\sum _{z=1/n}^{(n-1)/n} \to n \int _{0}^{1} dz$ and rewrite it as 
\begin{align}
\mathcal{S}_n & \simeq \frac{\sqrt{\pi}}{\sigma } n \int _{0}^{1} dz \frac{\Gamma\left( \frac{n(1-z)+1}{2}\right)}{\Gamma\left( \frac{n(1-z)+2}{2}\right)} ~\langle M _{n z} \rangle,~~~~\text{as }n \to \infty.  \label{largeSn-eq-2}
\end{align}
We next use the result of {\cite{Hartmann2020}} to write $\langle M _{n z} \rangle $ for large $n$ as $\langle M _{n z} \rangle \simeq \sigma \sqrt{\frac{2 z n}{\pi}} $. In addition, we approximate $\frac{\Gamma\left( \frac{n(1-z)+1}{2}\right)}{\Gamma\left( \frac{n(1-z)+2}{2}\right)} \simeq \sqrt{\frac{2}{n(1-z)}}$ as $n \to \infty$. Inserting these forms in Eq. \eqref{largeSn-eq-2} and performing the integration over $z$, we get
\begin{align}
\mathcal{S}_n \simeq \pi n, ~~~~~~~\text{as } n \to \infty.
\end{align}

\section{Derivation of $\langle M_s^2(n) \rangle$ in Eq. \eqref{fix-t-eq-14}}
\label{der-Ms2}
Here, we show that the expression of $\langle M_s^2(n) \rangle$ in Eq. \eqref{fix-t-eq-14} can be derived using the Pollaczek-Spitzer formula in Eq. \eqref{var-eq-1}. We first recall that $Q _s (M, n)$ in Eq.\eqref{fix-t-eq-12} represents the cumulative distribution that the maximum is less than $M$ up to $n$ steps for a random walker with independent and identically distributed  increments $\{ x_i \}$ chosen from the symmetric and continous distribution $g_s(x_i)$ in Eq. \eqref{fix-t-eq-7}. For this, the Pollaczek-Spitzer formula gives {\cite{Pollaczek1952, Spitzer1956}}  
\begin{align}
\sum _{n=0}^{\infty} z^n \langle e^{-\lambda M_s(n)} \rangle = \sum _{n=0}^{\infty} z^n \int_{0}^{\infty}d M e^{-\lambda M} Q_s'(M,n) = \frac{\phi_s (z, \lambda)}{\sqrt{1-z}},
\label{der-Ms2-eq-1}
\end{align}
where $0 \leq z \leq 1$ and $\lambda \geq 0$ and the function $\phi_s(z, \lambda)$ is defined as
\begin{align}
&\phi_s(z, \lambda) = \text{exp}\left(-\frac{\lambda}{\pi} \int_{0}^{\infty} d \xi ~\frac{\ln(1-z \hat{p}_s(\xi))}{\lambda ^2 +\xi^2} \right),~\text{ with}\label{der-Ms2-eq-2}\\
&\hat{p}_s(\xi) = \int _{-\infty}^{\infty}d x ~e^{i \xi x} g_s(x) ~= \frac{1}{\sqrt{1+\xi^2 \sigma _s ^2}}.
\label{der-Ms2-eq-3}
\end{align}
Here $\sigma _s = v_0 /(\gamma+s)$. As seen for fixed-$n$ ensemble in Eqs. \eqref{var-eq-6} and \eqref{var-eq-7}, one can  extend this formula to determine the generating function for the moments {\cite{Majumdar2010}}. For this case, one gets 
\begin{align}
&h^{(1)}_s(z) = \sum_{n=0}^{\infty} z^n \langle M_s(n)\rangle = \frac{1}{\pi (1-z)} \int_{0}^{\infty} \frac{d \xi}{\xi^2} ~\ln\left(\frac{1-z \hat{p}_s(\xi)}{1-z}\right), \label{der-Ms2-eq-4} \\
&h^{(2)}_s(z) = \sum_{n=0}^{\infty} z^n \langle M_s^2 (n)\rangle = (1-z) \left[h^{(1)}_s(z) \right]^2 +\frac{\sigma_s ^2 ~z}{2(1-z)^2}.
\label{der-Ms2-eq-5}
\end{align}
Taking derivative of $h^{(2)}_s(z)$ $n$-times, we get
\begin{align}
\langle M_s ^2 (n) \rangle = \sum _{m=1}^{n-1} \langle M_s(n) \rangle \left[ \langle M_{s}(n-m) \rangle-\langle M_{s}(n-m-1) \rangle\right] + \frac{n \sigma_s ^2}{2}.
\label{der-Ms2-eq-6}
\end{align}
We next use the results of {\cite{Hartmann2020}} to write $\langle  M_s(n) \rangle$ as
\begin{align}
\langle  M_s(n) \rangle = \frac{\sigma _s }{2\sqrt{\pi}} \sum _{j=1}^{n} \frac{\Gamma\left( \frac{j+1}{2}\right)}{\Gamma\left( \frac{j+2}{2}\right)}.
\label{der-Ms2-eq-7}
\end{align} 
and using this, we get
\begin{align}
\langle M_{s}(n-m) \rangle-\langle M_{s}(n-m-1) \rangle = \frac{\sigma _s }{2\sqrt{\pi}} ~\frac{\Gamma\left( \frac{n-m+1}{2}\right)}{\Gamma\left( \frac{n-m+2}{2}\right)}.
\label{der-Ms2-eq-8}
\end{align}
Finally, we insert Eqs. \eqref{der-Ms2-eq-7} and \eqref{der-Ms2-eq-8} in the expression of $\langle M_s ^2(n) \rangle $ in Eq. \eqref{der-Ms2-eq-6} and perform the sum over $m$ explicitly to yield
\begin{align}
\langle M_s^2(n) \rangle = \frac{\sigma _s ^2}{2} \left( \frac{\mathcal{S}_n}{\pi}+n \right),
\label{fix-n-eq-9}
\end{align}
where $\mathcal{S}_n$ is given in Eq. \eqref{fun-sn}. Identifying $\sigma _s = v_0 /(\gamma+s)$, we recover the result in Eq. \eqref{fix-t-eq-14}.

\end{document}